\documentclass[a4paper,twoside,final]{article}
\usepackage{amssymb,amsmath,amsfonts,amsthm}
\usepackage{graphicx,color,epsfig}
\usepackage{subfigure}
\usepackage{times}
\usepackage[
pdftitle={Anomalous scaling for 3d Cahn-Hilliard fronts},
pdfauthor={Timo Korvola, Antti Kupiainen, Jari Taskinen}
]{hyperref}

\newtheoremstyle{sftheorem}{}{}{\itshape}{}{\bfseries
  \sffamily}{.}{ }{}
\newtheoremstyle{sfdefinition}{}{}{\normalfont}{}{\bfseries
  \sffamily}{.}{ }{}
\newtheoremstyle{sfremark}{}{}{\normalfont}{}{\bfseries
  \sffamily}{.}{ }{\thmname{#1}\thmnote{ #3}}

\theoremstyle{sftheorem}
\newtheorem{theorem}{Theorem}
\newtheorem{proposition}{Proposition}
\newtheorem{lemma}{Lemma}
\newtheorem{corollary}{Corollary}
\newtheorem{conjecture}{Conjecture}

\theoremstyle{sfdefinition}
\newtheorem{definition}{Definition}
\newtheorem{example}{Example}

\theoremstyle{sfremark}
\newtheorem*{remark}{Remark}

\newcommand\ii{\mathrm i}           
\newcommand\bvec[1]{\mathrm{#1}}    
\newcommand\svec[1]{\mathbf{#1}}    
\newcommand\fvec[1]{\mathbf{#1}}    
\newcommand\dee{\mathrm d}          
\newcommand\deeC[1]{\frac{\dee #1}{2\pi\ii}} 
\newcommand\Lap{\mathop\triangle\nolimits} 
\newcommand\Oh{\mathcal O}          
\newcommand\zeroth[1]{\mathaccent23{#1}}
\newcommand\pole[1]{#1^{\textup{pole}}}
\newcommand\rest[1]{#1^{\textup{rest}}}
\newcommand\all[1]{#1^{\textup{all}}}
\DeclareMathOperator\Gtail{z}       

\let\Re=\undefined \DeclareMathOperator\Re{Re}
\let\Im=\undefined \DeclareMathOperator\Im{Im}

\newcommand\smartfrac[2]{
  \mathchoice{\textstyle\frac{#1}{#2}}{\frac{#1}{#2}}{
    \frac{#1}{#2}}{\frac{#1}{#2}}}
\newcommand{\hf}{{\smartfrac12}}
\newcommand{\td}{{\smartfrac13}}
\newcommand{\fh}{{\smartfrac14}}
\newcommand{\sh}{{\smartfrac16}}

\newcommand{\Nk}{\svec k}
\newcommand{\Np}{\svec p}
\newcommand{\Nx}{\svec x}

\newcommand\bbC{{\mathbb C}}

\newcommand\bbR{{\mathbb R}}
\newcommand\bbS{{\mathbb S}}

\newcommand\CB{{\mathcal B}}

\newcommand\cF{{\mathcal F}}
\newcommand\cL{{\mathcal L}}
\newcommand\CL{{\mathcal L}}

\newcommand{\p}{\partial}

\newcommand{\no}{\noindent}
\newcommand{\vs}{\vskip}
\newcommand\beq{\begin{equation}}
\newcommand\eeq{\end{equation}}
\newcommand\BET{\begin{theorem}}
\newcommand\ENT{\end{theorem}}
\newcommand\BEP{\begin{proposition}}
\newcommand\ENP{\end{proposition}}
\newcommand\BEL{\begin{lemma}}
\newcommand\ENL{\end{lemma}}
\newcommand\BEC{\begin{corollary}}
\newcommand\ENC{\end{corollary}}
\newcommand\BEE{\begin{example}}
\newcommand\ENE{\end{example}}
\newcommand\BER{\begin{remark}}
\newcommand\ENR{\end{remark}}
\newcommand\BED{\begin{definition}}
\newcommand\END{\end{definition}}
\newcommand\BECJ{\begin{conjecture}}
\newcommand\ENCJ{\end{conjecture}}

\title{Anomalous scaling for 3d Cahn--Hilliard fronts}
\author{Timo Korvola, Antti Kupiainen, Jari Taskinen}
\date{February 10, 2004}
\begin{document}
\maketitle

\begin{abstract}
  We prove the stability of the one dimensional kink solution
  of the  Cahn-Hilliard equation under $d$-dimensional perturbations
  for $d \geq 3$.
  We also establish a novel scaling behavior of the large time
  asymptotics of the solution. The leading asymptotics of the
  solution is characterized by a length scale proportional to $t^\td$
  instead of the usual $t^\hf$ scaling typical to parabolic problems.
\end{abstract}

\section{Introduction} 

The Cahn--Hilliard equation is a fourth order nonlinear evolution
equation for a real valued function $u(\bvec{x},t)$ defined on
some spatial domain $\bvec{x}\in\Omega\subset\bbR^d$:
\begin{align}
  \partial_t u &= \Lap ( -\Lap u - \hf u + \hf u^3 )  \label{CH} \\
  u( \bvec{x} ,0) &= g( \bvec{x}), \quad \bvec{x} \in \Omega.
  \label{initial}
\end{align}
The nonlinear term inside the brackets in the RHS has three
zeros, $u=0,\pm 1$ where the first is linearly unstable
whereas the others are linearly stable. 

The CH equation is used to model phase separation in mixtures
of two substances A,B (binary alloys) so that $u(\bvec{x},t)$
describes relative concentration of the substances
and the zeros $\pm 1$ correspond to pure phases
of A or B. 

When a random initial condition $g$ is given the solutions
of \eqref{CH} typically exhibit in numerical
simulations phase segregation, i.e.
domains of phase A and B start to form and increase in
size until they reach sizes comparable to the domain size.
To understand such extensive behavior of the solutions
it is natural to consider \eqref{CH} in the whole space 
$\Omega=\bbR^d$ which will be assumed in the present
paper. 

In one dimension a single phase boundary is described
by a stationary solution of \eqref{CH}, the so called
kink solution. This remains a solution also in 
dimensions $d>1$ and is given by
\begin{equation}
  u_0 (x) = \tanh (\hf x)
  \label{kink}
\end{equation}
where $x$ is the first coordinate of $\bvec{x}$.
Thus up to exponentially decaying tails, \eqref{kink}
describes a situation where we have phase B in the
domain $x<0$ and phase A in the
domain $x>0$. 

The presence of the fourth order derivative
in \eqref{CH} makes the mathematical analysis of 
the CH equation much harder than analogous
second order equations. The absence of a
spectral gap in unbounded  domains $\Omega$  due to the Laplacian
multiplying the RHS also complicates matters.
In \cite{bkt} and \cite{enza} the stability of the kink
solution in one dimensions was proved. Moreover in
\cite{bkt} the following leading asymptotics for $u(x,t)$
was established:
\[
u( x + a, t) = u_0( x) + \frac A{\sqrt{t}}\frac\dee{\dee{x}} \Bigl(
u_0(x)e^{-{\frac{x^2}{4t}}} \Bigr)
+ \frac B{\sqrt{t}}\frac\dee{\dee{x}}e^{-{\frac{x^2}{4t}}}+o(t^{-1})
\]
(in sup norm) where the constants $a,A,B$ depend on the
initial data and the function
\begin{equation}
  \frac\dee{\dee{x}} u_0(x)=(2\cosh^2(\hf x))^{-1}
  \label{dkink}
\end{equation}
decays as $2e^{-|x|}$. Thus, for large times one observes a
translated front (from the origin to $a$), a perturbation
of size $\Oh(\frac1{\sqrt{t}})$ localized near the origin and
a perturbation of size $\Oh(\frac1t)$ extending
to an interval of size $\sqrt{t}$ around the origin.
The latter exhibits typical diffusive scaling between space and time.

In the present paper we prove stability of the kink
solution when the spatial dimension $d\geq 3$ and establish
the following asymptotics for it. Let us agree to denote
variables in $\bbR^d$ by the letters $\bvec{x}$
or $\bvec{y}$ with $\bvec{x} = (x , \Nx ) \in 
\bbR \times \bbR^{d-1}$ and so on.
Moreover,  $\svec{k}$ or  $\svec{p}$ will only be in  $ \bbR^{d-1}$
with   $k := |\svec{k}|$ and the same for $p$.
Define the functions
\begin{equation}
  \phi^*(\Nx)=\int_{\bbR^{d-1}}e^{\ii\Nk\cdot\Nx}e^{-\td|\Nk|^3}
  \,\frac{\dee\Nk}{(2 \pi)^{d-1}}.
  \label{phi*}
\end{equation}
and for $t\geq 0$
\begin{equation}
  \phi(\Nx,t)=t^{-{\frac{d-1}{3}}}\phi^*({\frac{\Nx}{ t^\td}})
  \label{phi}
\end{equation}
or in terms of Fourier transform,
\begin{equation}
  \hat\phi(\Nk,t)=\hat\phi^*(t^\td\Nk)=e^{-\td t|\Nk|^3}.
  \label{phik}
\end{equation}
Let the initial datum \eqref{initial} be
given by
\begin{equation}
  g(\bvec{x})=u_0(x)+h(\bvec{x})
  \label{init1}
\end{equation}
and the function $h$ satisfy
\begin{equation}
  \Vert h\Vert_X:=\sup_\bvec{x}|h(\bvec{x})|(1+|\bvec{x}|)^r\leq \delta
  \label{init2}
\end{equation}
where $r>d+1$.
Then we prove

\BET \label{th} Let $d\geq 3$.
For $\delta$ small enough the equation \eqref{CH}
has a unique classical solution satisfying for $t\geq 1$
\begin{equation}
  u(\bvec{x},t)=u_0(x) +
  \frac A2 \partial_x u_0(x) \phi(\Nx,t)+\tilde u(\bvec{x},t)
  \label{asy}
\end{equation}
where
\begin{equation}
  \sup_\bvec{x}|\tilde u(\bvec{x},t)|\leq Ct^{-\frac{1}{12}}
  t^{-{\frac{d-1}{3}}}
  \label{asy1}
\end{equation}
and $A = \int_{\bbR^d} h( \bvec x) \,\dee\bvec x$.
\ENT

\begin{remark}[1]
  Since
  \[
  u_0(x) + \frac A2 \partial_x u_0(x) \phi(\Nx,t)
  = u_0(x + \frac A2 \phi(\Nx,t)) + \Oh(t^{-2{\frac{d-1}{3}}})
  \]
  we see that \eqref{asy} describes a a front that is translated in
  a domain of size $t^\td$ around the origin by a value of the order
  $t^{-{\frac{d-1}{3}}}$.  Note that in contrast to one dimension the
  translation of the front tends to zero as time tends to infinity.
  This is because a localized perturbation is not able to produce a
  constant shift in the whole transverse space $\bbR^{d-1}$.  However,
  the perturbation does not decay in the standard diffusive
  fashion but with the different power of time: $\sqrt{t}$ is replaced
  by $t^\td$. This scaling was argued to be present in the linearized
  CH equation in \cite{oono} and \cite{ro}.  We prove actually more
  detailed properties on the spatial behavior of $\tilde u$, see
  Proposition~\ref{prop1}.
\end{remark}
\begin{remark}[2]
  In two dimensions the nonlinearity becomes ``marginal'' in the
  terminology of \cite{bkl}. We do not know whether the asymptotics
  proved in the Theorem persists there.
\end{remark}

The remainder of the paper is organized as follows.  In
Section~\ref{linear} we present how the problem is reduced to a
nonlinear parabolic Cauchy problem with small initial data. We also
state the main estimates for its semigroup kernel needed for the
nonlinear analysis.  In Section~\ref{sec:proof-theorem} we use these
estimates to bound the nonlinear terms in the integral equation
corresponding to the above mentioned Cauchy problem, thus proving the
main result. The proofs for the crucial semigroup estimates are
presented in Sections \ref{sec:spectrum}--\ref{sec:sgsummary}.

\section{Linearization}\label{linear}

We start by separating the kink solution
\begin{equation}
  u = u_0 + v. \label{vee}
\end{equation}
Recalling that $u_0$ solves the CH equation we get for $v$ 
 the equation
\begin{equation}
  \partial_t v = \cL v
  + \Lap \bigl( \tfrac{3}{2} u_0 v^2 + \tfrac{1}{2} v^3 \bigr),
  \label{v1CH}
\end{equation}
where 
\begin{equation}
  \cL := - \Lap \bigl( \Lap +  \tfrac{1}{2} - \tfrac{3}{2} u_0^2 \bigr).
  \label{ell}
\end{equation}

This linear operator will play an important
role in the analysis since it will provide the leading asymptotics.
Indeed, we will solve the equation \eqref{v1CH} with
the initial condition $v_0=h$ by studying the
equivalent integral equation:

\begin{equation}
  v(t)  = e^{t \cL } h  + \int_0^t  e^{(t-s) \cL }
  \Lap \bigl( \tfrac32 u_0 v(s)^2 + \hf v(s)^3 \bigr) \,\dee{s}.
  \label{inteq}
\end{equation}

Let us therefore discuss the properties of the semigroup $\exp(t\CL)$
generated by $\CL$. Write the operator $\CL$ as
\begin{equation}
  \CL = \Lap H
  \quad \text{with} \quad
  H := -\Lap + 1 + V
  \quad \text{and} \quad
  V(x) := -\tfrac32 \cosh(\tfrac{x}2)^{-2}
  \label{ell1}
\end{equation}
Since $\CL$ is constant coefficient in the transverse $\Nx$
direction it will be convenient to work in a mixed
representation, with Fourier transform in these
variables. Thus given $f:\bbR^d \to \bbC$  denote
by $\hat f(x,\Nk)$  the Fourier transform 
with respect to the $d-1$ last coordinates. 
In this representation $-\CL$ becomes
\begin{equation}
  (-\Lap H f)\hat{}\;(x,\Nk)=(D_k H_k \hat f)( x,\Nk)
  \label{ellk}
\end{equation}
where 
\[
D_k = -\partial_{x}^2 + k^2, \hskip2cm
H_k = D_k + 1 + V
\]
and we denoted $|\Nk|$ by $k$.  From now on we will work in the
$(x,\Nk)$ representation and for notational simplicity drop the hats
from Fourier transforms.

The semigroup is then written as 
\begin{equation}
  (e^{t\CL}f)\;(x,\Nk)=\int_\bbR \dee y\,K(x,y,k,t) f(y,\Nk)
  \label{semi}
\end{equation}
with $K(x,y,k,t)$ the integral kernel of the semigroup
of the operator $-D_k H_k$. In this notation the integral
equation \eqref{inteq} becomes
\begin{equation}
  \begin{split}
    v(x , \Nk , t) = (e^{t \cL } h )( x , \Nk )
    + \int_0^t \int_\bbR
    &(\partial_{y}^2 - k^2) K( x, y, k, t -s) \\
    &\cdot \bigl( \tfrac32 u_0 (y) v^{*2}(y, \Nk, s)
    + \hf v^{*3}(y, \Nk, s) \bigr)
    \,\dee{s}\,\dee{y}
  \end{split}
  \label{inteq1}
\end{equation}
where the Laplacian was integrated by parts to act on
the semigroup kernel and $*$ denotes convolution
in the $\Nk$ variable.

We will express the semigroup as a
Dunford--Cauchy integral of the resolvent kernel:
\begin{equation}
  K(x,y,k,t)=\int_\Gamma \deeC{\zeta}\,
  e^{-\zeta t} (\zeta - D_k H_k)^{-1}( x, y)
  \label{reso}
\end{equation}
where $\Gamma$ is a suitable curve around the spectrum of $-D_k H_k$.

The resolvent kernel in \eqref{reso} may be studied by standard ODE
methods as was done in the one dimensional case in
\cite{bkt}. This is rather straightforward but tedious
and in this section we motivate and present a lemma
summarizing the estimates needed for the 
nonlinear analysis. The proof of the lemma is given
in Sections \ref{sec:spectrum}--\ref{sec:sgsummary}.

\begin{figure}
  \centering
  \epsfig{file=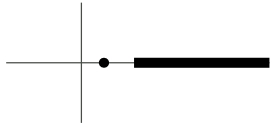}
  \caption{The spectrum of $D_k H_k$ (for small $k$).}
  \label{fig:spectrum}
\end{figure}

The spectrum of the operator $D_k H_k$ is on the positive real
axis. Furthermore, there exists a $k_0>0$ such that
for $k$ small, $k<k_0$, the spectrum contains
an isolated point
\begin{equation}
  \zeta_0=\td k^3+\Oh(k^4)
  \label{this-label-intentionally-left-blank}
\end{equation}
and the rest of the spectrum is on the semiaxis
$\bbS=[\frac34 k^2,\infty)$, see Figure~\ref{fig:spectrum}.
The resolvent has a simple pole at $\zeta_0$ and is analytic in the
complement of $\{\zeta_0\}\cup \bbS$. For $k$ larger than $k_0$ the
spectrum lies in $[b,\infty)$ with $b>0$.

Since the function $V$ in \eqref{ell1} decays exponentially
(as $-6e^{-|x|}$)
for large $x$ the behavior of the resolvent for
$x$ and $y$ large is determined by the functions
in the kernel
of the constant coefficient operator 
\[
\zeta - D_k (D_k + 1)
\]
obtained by setting $V$ to zero.  These are given by $e^{\mu x}$
with $\zeta -(-\mu^2+k^2) (-\mu^2+k^2 + 1)=0$ i.e.
\[
\mu = \pm\sqrt{\hf + k^2 \pm \hf \sqrt{1 + 4 \zeta}}.
\]
For large times $t$ the main contribution
to \eqref{reso} comes from 
small $\zeta$. In that domain the eigenvalues are approximately
\begin{equation}
  \mu_1\approx \sqrt{1+k^2+\zeta}
  \label{mu1}
\end{equation}
and
\begin{equation}
  \mu_2\approx\sqrt{k^2-\zeta}
  \label{mu2}
\end{equation}
and their negatives.

The integration contour in \eqref{reso} will be chosen
as follows. Let first $k\leq k_0$. Then

\vs 2mm

\begin{figure}
  \centering
  \subfigure[$k \le t^{-\hf}$]{\epsfig{file=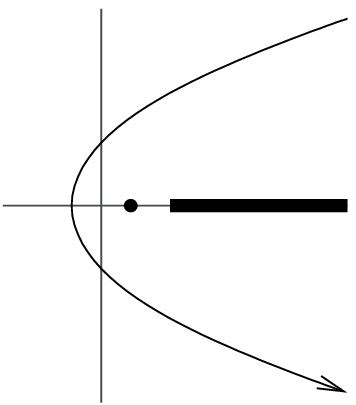}\label{fig:small-k}}
  \qquad
  \subfigure[$t^{-\hf} < k < k_0$]{\epsfig{file=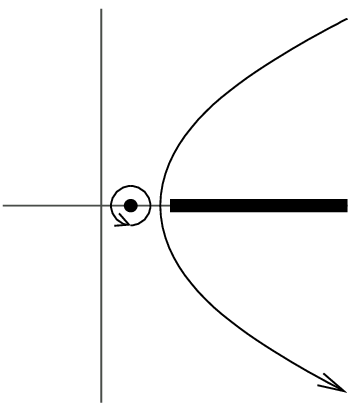}
    \label{fig:medium-k}}
  \caption{Integration paths for small values of $k$.}
\end{figure}

\no(a)  For $k\leq t^{-\hf}$ the contour is as in Figure~\ref{fig:small-k}.
In the neighborhood of the origin the eigenvalues
are $\mu_1\approx 1$ and $\Re\mu_2\approx ct^{-\hf}$
i.e. the decay of the resolvent is a combination
of 
\[
e^{-(1+\Oh(t^{-\hf}))|x|} \;\;{\rm and}\;\; e^{-\Oh(t^{-\hf})
{|x|}}
\]

\vs 2mm

\no(b)  For $ t^{-\hf}\leq k\leq k_0$ the contour
circles the pole and the semiaxis $\bbS$ as in Figure~\ref{fig:medium-k}.
At the pole the 
eigenvalues are 
\[
\mu_1=1+\Oh(k^2), \;\;\mu_2=(1+\Oh(k^2))k.
\]
On the second part of the contour and $|\zeta|\leq Ck^2$
$\mu_1$ is close to 1 and  $\mu_2$ has real part $\Oh(k)$ i.e.
larger than $t^{-\hf}$. 

\vs 2mm

The resolvent has a representation in terms
of the functions in the kernel of $\CL$ and its adjoint.
At $k=\zeta=0$ these are explicit and for small
$k$ and $\zeta$ they may be studied perturbatively. 
We will need explicitly a few of the leading
contributions to $K$ for small $k$. The following lemma
summarizes this.

\BEL \label{lm:K-smallk}
There exists $k_0>0$ such that for all $k<k_0$, $t\geq 1$ and
all $x,y$ the integral kernel $K(x,y,k,t)$ of the semigroup 
of the operator $-D_kH_k$ may be decomposed as
\begin{equation}
  K=K_0+K_1
  \label{Kdec}
\end{equation}
where
\begin{equation}
  K_0(x,y,k,t)= \hf \p u_0(x)Z(y,k,t)
  \label{k0}
\end{equation}
and 
\begin{equation}
  K_1(x , y , k ,t ) =
  \frac{1}{\sqrt{4\pi t}} \Bigl( e^{- \frac{(x-y)^2}{4t} } -
  e^{- \frac{(x + y)^2}{4t} }\Bigr)1_{xy>0} e^{-tk^2}
  +K_2(x , y , k ,t ).
  \label{timo2-}
\end{equation}
The function $Z$ is even in $y$ and has the property
\begin{equation}
  |\int_\bbR Z(y,k,t)\hat{h}(y,k)\dee{y}-e^{-\td tk^3}
  \int_{\bbR^d}h(\bvec{y})\dee\bvec{y}|
  \leq C\Vert h\Vert_X t^{-\td}e^{-\fh tk^3}
  \label{zprop}
\end{equation}
whereas 
\begin{multline}
  |K_2(x , y , k ,t )| \\
  \leq
  \begin{cases}
    Ct^{-1}e^{- c t^{-\hf} | x - y|} & \text{if $k \leq {t^{-\hf}}$,} \\
    C \bigl( t^{-1}e^{-\hf tk^2 - c k |x - y|} +k^{2} e^{-\fh tk^3 - ck
      | x - y| } \bigr) & \text{if $t^{-\hf} < k < {k_0}$.}
  \end{cases}
  \label{timo2--}
\end{multline}
\ENL
We will decompose in the same way also the kernel occurring
in eq.~\eqref{inteq1}:
\begin{equation}
  S:=(\p_y^2-k^2)K=S_0+S_1.
  \label{R}
\end{equation}
The following estimates hold:

\vs 2mm

\BEL \label{lm:S-smallk} {\rm (a)}
Let $k\leq t^{-1/2}$ and $t\geq 1$. Then
\begin{equation}
  |S_0(x , y , k ,t ) | \leq
  C(t^{-\hf}e^{-\hf( |x|+|y|)}
  +t^{-1} e^{- c (|x|+\frac{ | y|}{\sqrt{t} })})
  \label{timo1}
\end{equation}
and
\begin{equation}
  |S_1(x , y , k ,t ) |
  \leq C(t^{-1}e^{-\hf |x - y|} + t^{-1}e^{-c t^{-1/2} |x| -c |y|}
  + t^{-3/2}e^{- c \frac{|x- y|}{\sqrt{t} }}).
  \label{timo2}
\end{equation}

\no {\rm (b)}  Let $t^{-1/2}\leq k\leq k_0$ and $t\geq 1$. Then
\begin{equation}
  \begin{split}
    |S_0(x , y , k ,t )|
    &\leq C (ke^{-\hf( |x|+|y|)}
    + k^2 e^{-\hf | x|-ck|y| }) e^{-\fh tk^3 } \\
    &\quad+ C (t^{-\hf}e^{-\hf(|x|+|y|)}
    + t^{-1} e^{-\hf|x| - c k |y|}) e^{-\hf tk^2}
  \end{split}
  \label{timo3}
\end{equation}
and
\begin{equation}
  \begin{split}
    |S_1(x , y , k ,t ) |
    &\leq C ( k^{4} e^{-c k | x - y|} +  k^{3}e^{-ck|x|-c|y|}
    + k^{2} e^{-\hf |x - y|}) e^{-\fh tk^3} \\
    &\quad+ C (t^{-1} e^{-\hf |x-y|} + t^{-1} e^{-ck |x|-c|y|}
    + t^{-\frac32} e^{-c k |x- y| })e^{-\hf tk^2}.
  \end{split}
  \label{timo4}
\end{equation}
\ENL
\begin{remark}
  In order to get a feeling for the various terms, the following
  intuition is useful. For large $t$ and $x$ and for small $k$ the
  resolvent is built out of functions that decay approximately as
  $e^{-|x|}$, $e^{-k|x|}$ and $e^{-t^{-\hf}|x|}$.  Each $x$ derivative
  brings a factor of $1$, $k$ and $t^{-\hf}$ respectively. Thus, e.g.
  the third term in \eqref{timo2} behaves as the second derivate of
  $e^{t\Lap}$.
\end{remark}

Finally, the large $k$ or short time behavior of the semigroup is
dominated by the fourth derivatives in the symbol:

\BEL \label{bigk}{\rm (a)} Let $k> k_0$ and $t>1$. Then
\begin{equation}
  | K(x , y , k ,t ) |, \;\;| S(x , y , k ,t ) |  \leq Ce^{-\hf k^4t}
  e^{- c| x - y|}
  \label{l21}
\end{equation}
\no {\rm (b)} Let  $t<1$. Then for all $k$
\begin{equation}
  | K(x , y , k ,t ) |  \leq Ce^{-\hf k^4t}
  t^{-{\frac{1}{4}}} e^{- { {t}^{-\fh}| x - y|}}.
  \label{l22}
\end{equation}
and
\begin{equation}
  | S(x , y , k ,t ) |  \leq Ce^{-\hf k^4t}
  t^{-{\frac{3}{4}}} e^{- { {t}^{-\fh}| x - y|}}.
  \label{l23}
\end{equation}
\ENL

\section{Proof of the Theorem}
\label{sec:proof-theorem}

We solve \eqref{inteq1} using the contraction mapping principle in a
suitable Banach space. For each $t \geq 1$ we define the Banach space
$X_t$ of continuous functions $f:\bbR\times\bbR^{d-1} \to \bbC$ as
follows.  First, let $m=r-d+1$ ($r$ is defined in \eqref{init2} so
$m>2$) and define
\[
\omega (x ) := (1 + |x|)^{-m} .
\]
For $t\geq 1$ let
\begin{equation}
  k_t = \min\{k,1\}+\frac{1}{\sqrt{t}}
  \label{notat}
\end{equation}
and the same notation is used for any positive real number
in place of $k$.

The norm in $X_t$ is defined to be
\begin{equation}
  \Vert f \Vert_t := \sup\limits_{\substack{x \in \bbR \\ k \geq 0}}
  \bigl( \omega(x) + k_t\omega(k_tx) \bigr)^{-1}
  (1 + k^3 t)^{n} |f ( x , \Nk , t)|
  \label{Xnormi}
\end{equation}
Here $n$ can be taken arbitrary number larger than
$\frac{d+1}3$.\footnote{The third power comes from \eqref{zprop}.  The
  limit will provide sufficient $\svec k$-integrability for the proof
  of Lemma~\ref{lm:normconv}} %
The main estimate is the following
\BEP
\label{prop1} 
There exists a $\delta > 0$ such that 
if the  initial data satisfies $|h(\bvec{x})|\leq\delta
(1 + |\bvec{x}|)^{-r}$
then
the equation \eqref{v1CH} has
a unique classical solution $v$ such that, for all $t\geq 1 $,
$v(t) \in X_t$ and 
\begin{equation}
\Vert  v (t) - \frac A2 \partial_x u_0(x) \phi(\Nx,t) \Vert_t
\leq  C\delta t^{-{\frac{1 }{12}}}.
\label{tulos}
\end{equation}
\ENP

\no{\bf Remark.} Since for $\psi\in X_t$
\[
|\psi(\bvec{x})|\leq\int 
\dee\Nk( \omega (x) + k_t \omega (k_tx))
(1 + k^3 t)^{-n}\Vert\psi\Vert_t\leq Ct^{-{\frac{(d-1)}{3}}}\Vert\psi\Vert_t
\]
the sup norm of $v(t)$ is bounded by $Ct^{-{\frac{d-1}{3}}-\frac{1}{12} }$ and
the Theorem follows. 

The rest of this section is devoted to the proof of
Proposition~\ref{prop1}.  The proof splits into short times and long
times.  For short times we have the following lemma (see
\eqref{init2} to recall the definition of $\Vert \cdot \Vert_X$):
\BEL
\label{lemma3} 
There exists a $\delta > 0$ such that
if the  initial data satisfies $|h(\bvec{x})|\leq\delta
(1 + |\bvec{x}|)^{-r}$
then $v(1) \in X_1$,
\begin{equation}
  \Vert  v (1) \Vert_X \leq  C \delta
  \quad \text{and} \quad
  \Vert  v (1) \Vert_1  \leq  C \delta.
  \label{tulos1}
\end{equation}
\ENL

\begin{proof}
  This is quite standard: the leading symbol of the linearized
  equation is smoothing and preserves polynomial decay.  To prove the
  second estimate we need to control the large $k$ behavior, in view
  of the definition of the norm in \eqref{Xnormi}.  Here it is more
  natural to work in the $\bvec x$ representation and derive
  sufficient estimates for the derivatives.  Hence let $X^{(p)}$ be
  the space of $p$ times continuously differentiable functions $\bbR^d
  \to \bbC$ with the norm
  \[
  \|f\|_{X^{(p)}} := \max_{|\mu| \le p} \|\p_{\bvec x}^\mu f\|_X
  \]
  where $\mu$ is a multi-index.  We proceed to show that if the
  initial condition is in $X^{(p)}$ after an arbitrarily short time
  the solution will be in $X^{(p+1)}$.

  Write \eqref{v1CH} in the form
  \begin{equation}
    \partial_t v = -\Lap^2 v + \Lap N(v) 
    \label{ODY}
  \end{equation}
  with the initial condition $v(0) = h$ and
  \[
  N(v) := \hf \bigl( ( 3u_0^2 - 1) v +  3u_0 v^2 + v^3 \bigr).
  \]
  This is equivalent to the integral equation
  (after integration by parts)
  \begin{equation}
    v(t) = e^{-t \Lap^2} h +
    \int_0^t (\Lap e^{-(t-s) \Lap^2}) N(v(s)) \,\dee{s},
    \label{inteqshort}
  \end{equation}
  which can be differentiated:
  \begin{equation} \label{eq:dinteqshort}
    \p_{\bvec x}^\mu v( t) = e^{-t \Lap^2} \p_{\bvec x}^\mu h
    + \int_0^t \Lap e^{-(t-s) \Lap^2} \p_{\bvec x}^\mu N(v(s)) \,\dee s.
  \end{equation}
  From the explicit Fourier integral representation it is easy to see
  that the integral kernel $G(\bvec{x}, \bvec{y},t)$ of $e^{-t
    \Lap^2}$ satisfies the bound
  \[
  |\p_{\bvec y}^\nu G(\bvec{x}, \bvec{y},t)|
  \leq C_\nu t^{-\frac{d+|\nu|}4} e^{-t^{-1/4} |\bvec{x} - \bvec{y}|}
  \]
  for all multi-indices $\nu$ and $\bvec{x}, \bvec{y} \in \bbR^d$.
  Thus for any $f \in X$
  \[
  \begin{split}
    |(e^{-t \Lap^2} f)(\bvec x)|
    &\le \int_{\bbR^d} \frac C{t^{d/4}}
    e^{-\frac{|\bvec x - \bvec y|}{t^{1/4}}}
    \frac{\|f\|_X}{(1 + |\bvec y|)^r} \,\dee\bvec y \\
    &\le C \sup_{\bvec y} e^{-\frac{|\bvec x - \bvec y|}{2 t^{1/4}}}
    \frac{\|f\|_X}{(1 + |\bvec y|)^r}
    \int_{\bbR^d} \frac1{t^{d/4}}
    e^{-\frac{|\bvec x - \bvec y|}{2 t^{1/4}}} \,\dee\bvec y
    \le \frac{C \|f\|_X}{(1 + |\bvec x|)^r}
  \end{split}
  \]
  when $t \le 1$ and for any $g \in C([0,1], X)$
  \[
  \begin{split}
    \int_0^t \Lap e^{-(t-s) \Lap^2} g(s) \,\dee s
    &\le \frac C{(1 + |\bvec x|)^r}  \sup_{s \in [0,1]} \|g(s)\|_X
    \int_0^t \frac1{(t-s)^{1/2}} \,\dee s \\
    &\le \frac{C t^{1/2}}{(1 + |\bvec x|)^r}
    \sup_{s \in [0,1]} \|g(s)\|_X.
  \end{split}
  \]
  It follows that for any small enough $\tau > 0$ \eqref{inteqshort}
  can be solved by the contraction mapping principle in the Banach
  space $C([0,\tau],X^{(p)})$ with the maximum norm.  Namely, there
  exists $\delta_0>0$ s.t.\ if $\Vert h\Vert_{X^{(p)}} <\delta_0$ then
  $\Vert v \Vert_{C([0,\tau],X^{(p)})} \leq C \Vert h\Vert_{X^{(p)}}$.

  Differentiating \eqref{eq:dinteqshort} yet again yields
  \[
  \p_{\bvec x_j} \p_{\bvec x}^\mu v( t)
  = \p_{\bvec x_j} e^{-t \Lap^2} \p_{\bvec x}^\mu h
  + \int_0^t \p_{\bvec x_j} \Lap e^{-(t-s) \Lap^2}
  \p_{\bvec x}^\mu N(v(s)) \,\dee s.
  \]
  Estimate the integrals as in the previous case to get
  \[
  |\p_{\bvec x_j} \p_{\bvec x}^\mu v( t, {\bvec x})|
  < \frac C{(1 + |\bvec x|)^r} \Bigl(
  \frac1{t^{1/4}} \|\p_{\bvec x}^\mu h\|_X
  + t^{1/4} \sup_{\nu \le \mu} \|\p_{\bvec x}^\nu v\|_{C([0,\tau],X)}
  \Bigr).
  \]
  Thus the solution gains another derivative in an arbitrarily short
  time.  This can be iterated to get the estimate $\|v(1)\|_{X^{(3n)}}
  \le C \|h\|_X$, which proves the lemma.
\end{proof}

Let us denote
\[
f=v(1)
\]
and write our integral equation \eqref{inteq1} for $t\geq 1$
\begin{multline}
  v(x , \Nk , t) = (e^{(t-1) \cL } f )( x , \Nk ) \\
  + \int_1^t
  \int_\bbR S(x , y , \Nk ,t -s ) \, \bigl({\tfrac32}u_0 (y) v^{*2}
  (y , \Nk , s) + \hf v^{*3}(y , \Nk , s) \bigr) \dee{s} \dee{y}
  \label{inteq2}
\end{multline}
We want to prove that $v(t)$ satisfies the estimate \eqref{tulos}.
Let us
start with the solution of the linear equation:
\BEL \label{lemmajoku}
Let  $t \geq 1 $. The solution to the linearized problem is given by
\begin{equation}
  e^{(t-1) \cL } f = v_0+v_1
  \label{v01}
\end{equation}
\begin{equation}
  v_0(x,\Nk,t)=\frac A2 \p u_0(x)e^{-\td tk^3}
  \label{veenolla}
\end{equation}
with $A=\int_{\bbR^d} f$ and
\begin{equation}
  \Vert v_1\Vert_t\leq Ct^{-\td} (\Vert f\Vert_X + \Vert f\Vert_1).
  \label{v1}
\end{equation}
\ENL

\begin{proof}
  Let first $t \ge 2$ and $k \leq k_0$. Using \eqref{Kdec} decompose
  \[
  e^{(t-1) \cL } f =K(t-1)f=K_0(t-1)f+K_1(t-1)f:=w_0(t)+w_1(t).
  \]
  By \eqref{k0} and \eqref{zprop}
  \begin{equation}
    w_0(t,x,\Nk)=\frac A2 \p u_0(x)e^{-\td k^3 t} +\tilde w(x,\Nk,t)
    \label{aaa}
  \end{equation}
  with
  \[
  A=\int_{\bbR^d}f( \bvec x) \,\dee\bvec x
  \]
  and
  \begin{equation}
    |\tilde w(x,\Nk,t)|\leq C\Vert f\Vert_X t^{-\td}e^{-\fh k^3 t} e^{-|x|}
    \label{tildew}
  \end{equation}
  (we used $|\p u_0(x)|\leq 2e^{-|x|}$) whereby
  \[
  \Vert \tilde w\Vert_t \leq Ct^{-\td}\Vert f\Vert_X.
  \]

  For $w_1$ we use the decomposition of eq.~\eqref{timo2-} to write
  \[
  w_1=w_{11}+w_{12}.
  \]
  Start with the $w_{11}$ and let
  say $x\geq 0$. This is bounded by
  \begin{equation}
    |w_{11}(x,\Nk,t)|
    \leq \frac{e^{-tk^2}\Vert f\Vert_1}{\sqrt{4\pi t}} \int_0^\infty
    | e^{- \frac{(x-y)^2}{4t} } -
    e^{- \frac{(x + y)^2}{4t} }| (1+y)^{-m}\dee{y} .
    \label{diffe}
  \end{equation}
  Divide the integral to $y\leq \sqrt{t}$ and the complement.
  For the former, $I_1$, use
  \[
  | e^{- \frac{(x-y)^2}{4t} } -
  e^{- \frac{(x + y)^2}{4t} } |\leq e^{- \frac{x^2}{4t} }
  | e^{\frac{xy}{2t} } -
  e^{- \frac{xy}{2t} } |\leq {\frac{C}{\sqrt t}} e^{- \frac{x^2}{8t} }|y|
  \]
  to obtain
  \begin{align}
    I_1 &\leq C\Vert f\Vert_1 t^{-1}e^{- \frac{x^2}{8t} }e^{-tk^2}
    \leq C\Vert f\Vert_1 t^{-1}e^{-c(kx+{\frac{x^2}{ t}}) }e^{-\hf tk^2}
    \notag\\
    &\leq C\Vert f\Vert_1 t^{-\hf} k_t\omega(k_t x)e^{-\hf tk^2}
    \label{diffe0}
  \end{align}
  where ${\frac{x^2}{ t}}+tk^2\geq 2kx$ and $k_t^{-1}\leq \sqrt t$
  were used.
  The integral over $y\geq \sqrt{t}$ is bounded by
  \begin{equation}
    I_2= \frac{e^{-tk^2}\Vert f\Vert_1}{\sqrt{4\pi t}} \int_{\sqrt{t}}^\infty
    (e^{- \frac{(x-y)^2}{4t} } +
    e^{- \frac{(x + y)^2}{4t} } )(1+y)^{-m}\dee{y} .
    \label{diffe1}
  \end{equation}
  Consider the first term on the RHS. Integrate first over the
  domain $|x-y|>\hf x$. Since $(1+y)^{-m}\leq t^{-\hf}(1+y)^{-m+1}$
  and $m>2$ we may bound this term by
  \begin{equation}
    \frac{Ce^{-tk^2}\Vert f\Vert_1}{t}e^{- \frac{x^2}{16t} } 
  \end{equation}
  which can be absorbed to \eqref{diffe0}. The integral over
  $|x-y|\leq\hf x$ in turn is bounded by
  \begin{equation}
    C\Vert f\Vert_1 e^{-tk^2}(1+x)^{-m}.
    \label{diffe2}
  \end{equation}
  Since $x>{\tfrac23}\sqrt t$ in this domain we may bound \eqref{diffe2} as
  \begin{align}
    C\Vert f\Vert_1 e^{-tk^2}k_t^2(k_t+k_t x)^{-2}(1+x)^{-m+2}
    &\leq C\Vert f\Vert_1 e^{-tk^2}k_t^2 (1+k_tx)^{-m}
    \notag\\
    &\leq C\Vert f\Vert_1 t^{-\hf}k_te^{-\hf tk^2} (1+k_tx)^{-m}.
    \label{diffe3}
  \end{align}
  The second term on the RHS of  \eqref{diffe1} is bounded
  by the first one.  Thus, altogether, we got
  \begin{equation}
    |w_{11}|\leq  Ct^{-\td}(\omega(x)+k_t\omega(k_tx))e^{-\fh tk^3}.
    \label{w11}
  \end{equation}

  For $w_{12}$ we use \eqref{timo2--}. This estimate, for $k\leq t^{-\hf}$,
  is readily seen to produce a bound like \eqref{w11} and, for $k_0\geq
  k\geq t^{-\hf}$ similarly, 
  if we use $ke^{-\fh tk^3}\leq Ct^{-\td}e^{-\sh tk^3}$
  (and replace in \eqref{w11} the $\fh$ by $\sh$).

  Thus, altogether, in the domain $k\leq k_0$, we have proved
  the decomposition \eqref{v01} with 
  \begin{equation}
    |v_1(x,\Nk,t)|
    \leq Ct^{-\td}(\omega(x)+k_t\omega(k_tx))e^{-\sh tk^3}.
    \label{v1f}
  \end{equation}

  For $t \ge 2$ and $k>k_0$ \eqref{l21} gives
  \begin{equation}
    \vert (e^{(t-1) \cL } f)(x,\Nk) \vert
    \leq C \|f\|_1 e^{-\hf k^4 (t-1)}\omega(x)
    \leq C \|f\|_1 e^{-ct}e^{-\fh (t-1) k^3}\omega(x) \label{l43}
  \end{equation}
  upon using $\int e^{-|x-y|}\omega(y)\dee{y}\leq C\omega(x)$.
  Since
  \[
  \Vert A \p u_0 e^{-\td t k^3}1_{k>k_0}\Vert_t\leq \|f\| e^{-ct}
  \]
  it can be absorbed into \eqref{v1}.

  For $t \le 2$ use \eqref{l22} and \eqref{e2}:
  \[
  |(e^{(t-1) \cL} f)(x)|
  \le \frac{C \|f\|_1}{(1 + k^3)^n (t-1)^{1/4}}
  \int_\bbR e^{-\frac{|x - y|}{(t-1)^{1/4}}} \omega( y) \,\dee y
  \le \frac{C \|f\|_1 \omega( x)}{(1 + k^3)^n}.
  \]
  Again this and $v_0$ can be absorbed into $v_1$.
\end{proof}

Define next $w=v-e^{(t-1) \cL}$ and $w_1=w+v_1$, i.e.,
\[
v = v_0 + w_1 = v_0 + v_1 + w.
\]
We show that $w$ and $w_1$ remain bounded in the Banach space $\CB$ of
continuous functions $s\to w(s)\in X_s$ with norm
\begin{equation}
  \Vert w\Vert:=\sup_{s\geq 1}s^{\frac{1}{12}} \Vert w(\cdot,\cdot,s)\Vert_s.
  \label{essresult}
\end{equation}
Proposition~\ref{prop1} follows since $v = v_0 + w_1$.

Assume now that $t > 2$.
Since the estimates for our semigroup are quite different for short and
long times, we decompose the integral equation as
\begin{align}
  w(t) &= \int_1^{t-1} S(t - s)
  ({\tfrac32}u_0  (v_0(s)+w_1(s))^{*2} + \hf v(s)^{*3}) \,\dee{s}
  + C(t)
  \label{inteq3}\\
  C(t) &= \int_{t-1}^t S(t - s)
  ({\tfrac32}u_0v(s)^{*2} + \hf v(s)^{*3}) \,\dee{s};
  \label{inteq30}
\end{align}
for later convenience we wrote in the first term $v=v_0+w_1$.
(We omit the details of the  case $ 1 < t \leq 2$. This is  simpler:
one omits the term with $\int_1^{t-1} $ on the
right hand side of \eqref{inteq3},
and the integral in \eqref{inteq30} is replaced by $\int_1^t$. One applies 
the estimates given for $C(t)$ below.)

We assume that \eqref{tulos1} holds and that
$w$ is in the ball with radius $\delta$ at
the origin of $\CB$ and show the RHS of \eqref{inteq3} 
stays in this ball and contracts there. Then \eqref{tulos} follows
from the contraction mapping principle.
Notice that $v$ satisfies by \eqref{veenolla}, \eqref{v1}
and \eqref{essresult}
\begin{equation}
  \sup_{s\leq t} \Vert v(\cdot,\cdot,s)\Vert_s
  \leq C\delta
  \label{vbound}
\end{equation}
since $|A|\leq C\delta$.

Next, recall that for $t-s>1$ we have the decomposition 
$S=S_0+S_1$ where the operator $S_0$
annihilates odd functions since $Z$ in \eqref{k0} is even. 
Since $u_0$ is odd and $v_0$
even, the term involving $S_0u_0v_0^{*2}$ vanishes and 
we may rewrite \eqref{inteq3} as 
\begin{align}
  w(t) &= \int_1^{t-1} S_0(t -s )
  ({\tfrac32}u_0(v_0+w_1)*w_1 + \hf v^{*3} )\,\dee{s}
  \notag\\
  &\quad+ \int_1^{t-1} S_1(t -s )
  ({\tfrac32}u_0 v^{*2} + \hf v^{*3})\,\dee{s} + C(t)
  \notag\\
  &=: A(t)+B(t)+C(t).
 \label{inteq4}
\end{align}
We need a lemma about how our norm behaves under convolutions:
\BEL \label{lm:normconv}
Let $f,g\in X_s$. Then
\begin{equation}
  \Vert f*g\Vert_s \leq Cs^{-{\frac{d-1}{3}}}
  \Vert f\Vert_s\Vert g\Vert_s
  \label{ba}
\end{equation}
\ENL
\begin{proof}
  We get from the definition of the norm
  (using the notation of \eqref{notat} )
  \begin{equation}
    \begin{split}
      |(f*g)(x,\Nk)| \leq \Vert f\Vert_s \Vert g\Vert_s \int_{\bbR^{d-1}}
      &\bigl( \omega(x) + |\Nk-\Np|_s \omega(|\Nk-\Np |_sx) \bigr) \\
      &\cdot \bigl( \omega(x)+p_s \omega(p_s x) \bigr) \\
      &\cdot (1+|\Nk-\Np |^3s)^{-n}(1+p^3s)^{-n} \,\dee\Np.
    \end{split}
    \label{a0}
  \end{equation}
  Expanding the product the integral gives rise to four contributions.
  Three of these have at least one factor $\omega(x)$
  and can be estimated by  $C \omega (x) I$, where
  \begin{equation}
    I := \int (1+|\Nk-\Np |^3s)^{-n}(1+p^3s)^{-n}\dee\Np.
    \label{a011}
  \end{equation}
  Dividing the integration domain to $E:= \{ |\Np - \Nk |\leq \hf k \}$ and
  the complement $E_c$, we estimate
  \begin{align}
    I &\leq (1+(\hf k)^3s)^{-n}
    \Bigl( \int_{ E } (1+|\Nk - \Np |^3s)^{-n} \,\dee \Np
    + \int_{ E_c } (1+p^3s)^{-n} \,\dee\Np \Bigr)
    \notag\\
    &\leq C (1+ k^3s)^{-n} \int_{\bbR^{d-1} }(1+p^3s)^{-n} \,\dee\Np
    \leq C (1+k^3s)^{-n}s^{-{\frac{d-1}{3}}}.
    \label{convo}
  \end{align}
  The remaining term $I'$ is more complicated:
  \begin{align*}
    I' &:= \int p_s|\Nk-\Np |_s\omega(p_sx) \omega(|\Nk-\Np |_sx)
    (1+|\Nk-\Np |^3s)^{-n}(1+p^3s)^{-n} \,\dee\Np \\
    &\leq C k_s \omega(k_s x) (1+k^3s)^{-n} \int_E
    \omega(|\Nk-\Np |_sx)(1+|\Nk-\Np |^3s)^{-n} \,\dee\Np \\
    &\quad+ C \omega(k_s x) (1+k^3s)^{-n} \int_{E_c}
    p_s |\Nk-\Np |_s \omega(p_s x) (1+p^3s)^{-n}\dee\Np.
  \end{align*}
  The first term is of the appropriate form, since the integral is
  bounded by $Cs^{-(d-1)/3}$. We need to extract a $k_s$ also from the
  second integral.  Use $|\Nk - \Np|_s \leq k_s + p_s$. The term
  containing $k_s$ is clear, and it remains to estimate
  \begin{align*}
    \int_{E_c} p_s^2 \omega(p_sx) (1+p^3s)^{-n} \dee\Np
    &\leq \int_{\bbR^{d-1} } \Bigl( p + \frac{1}{\sqrt{s}} \Bigr)^2
    (1+p^3s)^{-n} \dee\Np \\
    &\leq \frac{C}{s^{(d+1)/3} } \leq \frac{Ck_s}{s^{(d-1)/3}}.
  \end{align*}
\end{proof}

Let us start bounding the terms in \eqref{inteq4} and consider
first $C(t)$.
By Lemma~\ref{lm:normconv} and \eqref{vbound} we have
\begin{equation}
  \Vert v^{*2}
  \Vert_s  \; , \;\Vert v^{*3}
  \Vert_s \leq C\delta^2s^{-{\frac{d-1}{3}}}.
  \label{esti}
\end{equation}
Thus 
the bound \eqref{l23} gives 
\begin{multline}
  |C(t)|
  \leq C\delta^2
  \int_{t-1}^t 
  \dee{s}\, e^{-\hf k^4(t-s)}
  (t-s)^{-{\frac{3}{4}}}
  \int\dee{y}\, e^{- { {(t-s)}^{-\fh}| x - y|}} \\
  \cdot
  (\omega(y)+k_{s}\omega(k_{s}y))(1+sk^3)^{-n}s^{-{\frac{d-1}{3}}}.
  \label{BBB}
\end{multline}
The following simple estimate is needed repeatedly below:
\BEL \label{lm:expconv}
Let $k,p>0$.  Then
\begin{equation}
  \int \dee y e^{- k|x- y|}
  \omega(py)\leq\begin{cases}
    Ck^{-1}\omega(px) & \text{\ for $k>p$,}\\
    C(p^{-1}\omega(kx)+k^{-1}\omega(px)) & \text{\ for $k\leq p$.}
  \end{cases}
  \label{e2}
\end{equation}
\ENL
\begin{proof}
 For $k>p$ we decompose the integral to the set $E$
given by $|x-y| \le \hf|x|$ and its complement $E_c$:
\[
\int e^{-k|x-y|} \omega( p y) \,\dee y
\le \omega( \hf p x) \int_E e^{-k |x - y|} \,\dee y
+ e^{-\fh p|x|} \int_{E_c} e^{- \hf k|x- y|} \omega(py) \,\dee y
\]
and \eqref{e2} follows.
\end{proof}

Using Lemma~\ref{lm:expconv} and
$e^{-(t-s)^{-1/4}|x-y|} \leq  e^{-|x-y|} $
we may now bound the $y$ integral in \eqref{BBB}
by
\[
\int \dee{y} e^{-| x - y|}(\omega(y)+k_{s}\omega(k_{s}y))
\leq C(\omega(x)+k_{s}\omega(k_{s}x))
\]
Since $s\in[t-1,t]$, by changing the constant $C$ we may replace
$s$ by $t$ except in the $t-s$ factors and end up with
\[
|C(t)| \leq C
\delta^2t^{-{\frac{d-1}{3}}}
(\omega(x)+k_t\omega(k_tx))(1+tk^3)^{-n}
\]
i.e. the $\CB_t$ norm 
is bounded by 
\begin{equation}
  \Vert C(t)\Vert_t\leq C\delta^2t^{-{\frac{d-1}{3}}}
  \label{B1}
\end{equation}

For $A$ and $B$ in \eqref{inteq4} we need to
distinguish between the various $k$ values.

\vs 2mm

\no (a). Let first $k\leq t^{-\hf}$. Using \eqref{esti} and \eqref{timo2} 
we bound  $B(t)$
by
\begin{align}
  |B(t)| &\leq C\delta^2 (I_1 + I_2 + I_3)
  \notag\\
  I_1 &:= \int_1^{t-1} \dee{s}\int \dee{y}\,
  (t-s)^{-1}e^{-\hf |x - y|}s^{-{\frac{d-1}{3}}}
  (\omega(y) + k_s\omega(k_s y))
  \label{I1}\\
  I_2 &:= \int_1^{t-1} \dee{s}\int \dee{y}\,
  (t-s)^{-1}e^{- c \frac{|x|}{\sqrt{t-s} } - c |y|}
  s^{-{\frac{d-1}{3}}} ( \omega(y) + k_s\omega(k_sy))
  \label{I2}\\
  I_3 &:= \int_1^{t-1} \dee{s}\int \dee{y}\,
  (t-s)^{-3/2}e^{-c \frac{|x - y|}{\sqrt{t-s}}}
  s^{-{\frac{d-1}{3}}} ( \omega(y) + k_s\omega(k_sy))
  \label{I3}
\end{align}
For $I_1$ we use Lemma~\ref{lm:expconv}, $s^{-1/2} < k_s < 2s^{-1/2}$
and $\omega(x / \sqrt{s}) \leq\omega(x / \sqrt{t})$ to deduce
\begin{align}
  I_1 &\leq C \int_1^{t-1} \dee{s}\, (t-s)^{-1} s^{-{\frac{d-1}{3}}}
  \Bigl( \omega(x) + \frac{1}{\sqrt{s}} \omega(\frac{x}{\sqrt{s}} ) \Bigr)
  \notag\\
  &\leq \frac{C\log t}{t^{{\frac{1}{3}}} }
  \Bigl( \omega(x) + k_t \omega(k_t x ) \Bigr).
  \label{I11}
\end{align}
For $I_2$ we use simply
\[
I_2 \leq \int_1^{t-1}  \frac{\dee{s} } {(t-s) s^{{\frac{d-1}{3}}} }
\omega(\frac{x}{\sqrt{t}} ) \leq  \frac{C\log t}{t^{{\frac{d-1}{3}}} }
\omega(\frac{x}{\sqrt{t}} ),
\]
assuming $d \leq 4$. (The bound is not useful in the 
case $d=2$.)  For all $d \geq 3$ we obtain
$I_2 \leq C t^{-1/6} \log t \  k_t \omega(k_t x )$.

For $I_3$ we use Lemma~\ref{lm:expconv} 
again, then estimate both $\omega(x / \sqrt{s}) $ and
$\omega(x / \sqrt{t- s})$  by $\omega(x / \sqrt{t})$:
\begin{align}
  I_3 &\leq C \int_1^{t-1} \dee{s}\, s^{-{\frac{d-1}{3}}}
  (t-s)^{-3/2} \Bigl( \omega(\frac x{\sqrt{t-s}}) +\sqrt{t-s}\omega(x)
  +\sqrt{\frac{t-s}{s}}\omega(\frac x{\sqrt{s}} ) \Bigr)
  \notag\\
  &\leq \frac{C}{t^{\frac{d-1}{3}} } \Bigl( \omega(x) \log t +
  \omega(\frac{x}{\sqrt{t}} ) \Bigr) + \frac{C \log t }{ t^{5/6} }
  \omega(\frac{x}{\sqrt{t}}) )
\label{b20}
\end{align}
for $d\leq 4$. Again for $d=2$ the result is not useful.
For $d > 4$ one obtains $I_3 \leq C t^{-1}
(\omega(x) + \omega(x/\sqrt{t} ) ) $.
Thus for $d \geq 3$, 
\begin{equation}
  |B(t)|\leq C\delta^2 t^{-1/6} \log t \;
  \Bigl( \omega(x) + t^{-\hf}\omega(\frac{x}{\sqrt{t}}) \Bigr).
  \label{b2}
\end{equation}
For the $A(t)$ in \eqref{inteq4} we use
Lemma~\ref{lm:normconv} to bound
\[
\Vert {\tfrac32}u_0(v_0+w_1)*w_1
+ \hf v^{*3}\Vert_s\leq C\delta^2(s^{-{\frac{d-1}{3}}-\frac{1}{12}}+
s^{-2{\frac{d-1}{3}}})\leq  C'\delta^2 s^{-{\frac{d-1}{3}}-\frac{1}{12}}
\]
and then \eqref{timo1} to get the bound
\[
\begin{split}
  C \delta^2 \int_1^{t-1} \dee{s}\,
  &\bigl( (t-s)^{-\hf} e^{-\hf (|x|+|y|)}
  + (t-s)^{-1} e^{- c ( |x|+(t-s)^{-\hf}|y|)} \bigr) \\
  &\cdot s^{-{\frac{d-1}{3}}-\frac{1}{12}}( \omega(y) + k_s\omega(k_sy))
\end{split}
\]
which for $d\geq 3$  is bounded by
\begin{equation}
  C\delta^2 t^{-\sh}\omega(x).
  \label{b3}
\end{equation}
Thus for $k\leq t^{-\hf}$
\begin{equation}
  |A(t)+B(t)|\leq C\delta^2 \log t \; t^{-1/6}
  \Bigl( \omega(x) + t^{-\hf}\omega(\frac{x}{\sqrt{t}}) \Bigr).
  \label{B2}
\end{equation}

\vs 2mm

\no (b)Let us next consider $k\in[t^{-\hf},k_0]$ and start again with the
 $B(t)$ in \eqref{inteq4}. Using \eqref{timo4} $|B(t)|$ is bounded 
by the sum of
\begin{multline}
  C \delta^2 \int\limits_1^{t-1} \dee s \int \dee y\,
  \bigl( (t-s)^{-1}e^{-\hf |x - y|} + (t-s)^{-1}e^{- c k|x|- c|y|}
  + (t-s)^{-3/2}e^{- c k|x- y|} \bigr) \\
  \cdot s^{-{\frac{d-1}{3}}} \bigl( \omega(y) + k_s\omega(k_sy) \bigr)
  e^{-\hf (t-s)k^2 }(1+sk^3)^{-n}
  \label{int2}
\end{multline}
and
\begin{multline}
  C \delta^2 \int_1^{t-1} \dee{s} \int \dee{y}\,
  \bigl( k^{4} e^{- c k | x - y| } + k^3 e^{-c (k|x|+|y|)}
  + k^2 e^{-\hf |x- y|} \bigr) \\
  \cdot s^{-{\frac{d-1}{3}}}
  \bigl( \omega(y) + k_s\omega(k_sy) \bigr)
  e^{-\fh (t-s)k^3} (1+sk^3)^{-n}.
  \label{int3}
\end{multline}
The first integral resembles the case $k < t^{-1/2}$. 
To estimate \eqref{int2} use
\[
e^{-\hf (t-s)k^2}(1+sk^3)^{-n}\leq C(1+tk^3)^{-n},
\]
and assume $d \leq 4$. Then 
\eqref{int2} is bounded by
\begin{multline}
  C \delta^2 (1+tk^3)^{-n} \int\limits_1^{t-1} \dee{s}\,
  \frac{1}{s^{{\frac{d-1}{3}}} (t-s)}
  \Bigl( \omega(x) + \omega(kx) + \frac{1}{k \sqrt{t-s} }
  ( \omega(x) + k_s\omega(k_s x)) \Bigr) \\
  \leq C \delta^2 (1+tk^3)^{-n}
  \Bigl( \frac{\log t}{t^{{\frac{d-1}{3}}}} (\omega(x) + \omega(k_tx))
  + \frac{1}{t^{{\frac{d-1}{3}}}} ( \sqrt{t} \omega(x) + \omega(k_t x))
  \Bigr).
  \label{int21}
\end{multline}
In the case $d > 4 $ replace $(d-1)/3$ by $1$.
For $d \geq 3$ we thus get the bound
\begin{equation}
  C \delta^2 (1+tk^3)^{-n}
  \frac{\log t}{t^{1/6}}
  (\omega(x) + k_t\omega(k_tx)) .
  \label{int2b}
\end{equation}

Thus \eqref{int3} remains. Using \eqref{e2} we have
\[
\int e^{-\hf k | x - y| }\omega(y)\dee{y}\leq
C( k^{-1}\omega(x) + \omega(kx))
\]
and 
\[
\int e^{-\hf k | x - y| }k_s\omega(k_sy)\dee{y}\leq
C( k^{-1}k_s\omega(k_sx) + \omega(kx))
\]
whereby
\begin{multline*}
  \int\dee{y}\,
  \bigl( k^{4} e^{-c k |x - y|} + k^3e^{-c (k|x|+|y|)} +
  k^2e^{-\hf |x - y|} \bigr)
  \bigl( \omega(y) + k_s\omega(k_s y) \bigr) \\
  \leq Ck^{2} \bigl( \omega(x)+k_s\omega(k_sx) + k\omega(kx) \bigr)
  \leq Ck^{2} \bigl( \omega(x)+(k_t+s^{-\hf})\omega(k_tx) \bigr)
\end{multline*}
since $\omega(kx)$ and $\omega(k_sx)$ are bounded
by $C\omega(k_tx)$ for $k>t^{-1/2}$.
Thus 
\[
\eqref{int3}
\leq C\delta^2 \int_1^{t-1} \dee{s}\, s^{-{\frac{d-1}{3}}}
e^{-\fh (t-s)k^3 }(1+sk^3)^{-n} k^{2}
\bigl( \omega(x)+(k_t+s^{-\hf})\omega(k_tx) \bigr).
\]
Using
\[
e^{-\fh (t-s)k^3 }(1+sk^3)^{-n}k^2\leq C(t-s)^{-2/3}
(1+tk^3)^{-n}
\]
we get
\[
\eqref{int3}
\leq C\delta^2 (\omega(x)+k_t\omega(k_tx))(1+tk^3)^{-n}
\int_1^{t-1} \dee{s}\,
s^{-{\frac{d-1}{3}}}(t-s)^{-2/3} \Bigl( 1+({\frac{t}{ s}})^\hf \Bigr).
\]
The integral is, for $d=3$, bounded by $Ct^{-\sh}$ and
we end up with the bound for $k\in[t^{-\hf},k_0]$
\begin{equation}
  \eqref{int3}
  \leq C\delta^2 t^{-\sh}(\omega(x)+k_t\omega(k_tx))(1+tk^3)^{-n}
  \label{b40}
\end{equation}

$A(t)$ in \eqref{inteq4} is for $k\in[t^{-\hf},k_0]$
bounded by the sum of
\begin{multline}
  C\delta^2\int \dee{y}\int_1^{t-1} \dee{s}\,
  \bigl( k^{2} e^{-\hf | x| - ck|y|}
  + k e^{-\hf(|x|+|y|)} \bigr) \\
  \cdot s^{-{\frac{d-1}{3}}}s^{-1/12}(\omega(y) + k_s\omega(k_sy))
  e^{-\fh (t-s)k^3 }(1+sk^3)^{-n}
  \label{worst}
\end{multline}
and 
\begin{multline*}
  C \delta^2 \int\dee{y} \int_1^{t-1} \dee{s}\,
  \bigl( (t-s)^{-\hf}e^{-\hf( |x|+|y|)}
  + (t-s)^{-1} e^{- \hf|x| - c k | y|} \bigr) \\
  \cdot s^{-{\frac{d-1}{3}}} s^{-1/12}
  (\omega(y) + k_s\omega(k_sy))e^{-\hf (t-s)k^2 }(1+sk^3)^{-n}
\end{multline*}
In the former expression we replace as above $k$ by $(t-s)^{-1/3}$ and
end up with the bound
\begin{equation}
  C\delta^3 t^{-1/12}\omega(x)(1+tk^3)^{-n}.
  \label{b4}
\end{equation}
and the latter is smaller with $t^{-1/12}$ replaced by $t^{-1/6}$.
Thus for $k\in[t^{-\hf},k_0]$
\begin{equation}
  |A(t)+B(t)|\leq C\delta^3 t^{-1/12}
  (\omega(x)+k_t\omega(k_tx))(1+tk^3)^{-n} .
  \label{B3}
\end{equation}

\vs 2mm

\no (c)
Finally we suppose $k>k_0$ and use the bound \eqref{l21}
and $\omega(k_sx)\leq C\omega(x)$ which holds for such $k$
to get
\begin{align}
  |A(t)+B(t)|
  &\leq C\delta^2 \int_1^{t-1} \dee{s} \int\dee{y}\,
  e^{-\hf k^4(t-s)} e^{- c| x - y|}
  s^{-{\frac{d-1}{3}}} \omega(y)(1+s k^3)^{-n}
  \notag\\
  &\leq C\delta^2 t^{-2/3}\omega(x)(1+tk^3)^{-n}
  \label{B4}
\end{align}
The bounds \eqref{B1}, \eqref{B2}, \eqref{B3} and \eqref{B4}
show that our ball is mapped to itself. The contractive
property goes similarly.

\section{The spectrum}
\label{sec:spectrum}

In the remaining sections we will present the linear analysis needed
for the semigroup kernel estimates in Lemmas \ref{lm:K-smallk},
\ref{lm:S-smallk} and \ref{bigk} of Section~\ref{linear}.  Since the
linearized Cahn--Hilliard operator is of the fourth order, the
calculations are lengthy, and we will be brief in the most tedious
details.  The full details may be found in the thesis \cite{timo}.
The linear analysis is based on an approach outlined in
\cite{PegoWeinstein}.

We remark that, in addition to the existing notations, boldface
letters will mainly refer to vectors in $\bbC^4$ and $4 \times 4$
matrices in the sequel. A boldface symbol and the corresponding italic
symbol are not always mathematically related: consider for example
$\fvec R$ and $R$, or $\fvec v$ and $v$.

The following lemma summarizes the properties of
the spectrum of $D_k H_k$ alluded to in Section~\ref{linear}.
\begin{lemma} \label{lm:coarse-spectrum}
  For any $k > 0$ the spectrum of $D_k H_k$ is contained in $[k^4,
  \infty)$.  There exists an $\epsilon > 0$ such that for $0 < k <
  \epsilon$ the bottom of the spectrum is an isolated eigenvalue
  $\zeta_0$ of multiplicity one with $\zeta_0 \le C k^3$.  The
  remaining part of the spectrum is bounded from below by ${\frac{3 }{4}}
  k^2$.
\end{lemma}
\begin{proof}
  $D_k H_k$ has asymptotically constant coefficients, thus the
  essential spectrum is $[k^2+k^4,\infty)$, the same as that of the
  constant coefficient limit $D_k^2+ D_k$, see
  \cite[Proposition~26.2]{ColletEckmann:IFES}.  Eigenvalues on the
  other hand are the same as those of the self-adjoint operator $A_k
  := D_k^{1/2} H_k D_k^{1/2}$: the eigenvectors are mapped onto each
  other by $D_k^{1/2}$, which is invertible when $k > 0$: we have the
  convolution kernel $(D_k^{-1})(x) =: G_k(x) = (2k)^{-1} e^{-k|x|}$.
  
  $H_0$ has two isolated eigenvalues at $0$ and $3/4$ and a continuous
  spectrum $[1,\infty)$, see \cite[page 79]{LandauLifshitz:QM}.  The
  eigenfunctions are $V$ from \eqref{ell1} and $x \mapsto \sinh(x/2) /
  \cosh(x/2)^2$.

  The spectrum of $A_k$ can be studied with the minimax principle:
  \[
  \zeta_0 = \inf_u \frac{\langle u, H_k u \rangle}{\langle u, G_k u \rangle}
  \le \frac{k^2}{\langle V, G_k V \rangle} \le C k^3
  \]
  for small $k$.  We also get the lower bound:
  \[
  \zeta_0 \ge \inf_u \frac{k^4\|u\|^2}{\langle u, k^2 G_k u \rangle}
  \ge k^4.
  \]
  To see that $\zeta_0$ is a discrete eigenvalue of multiplicity one
  we proceed further with the minimax principle:
  \begin{align*}
    \zeta_1
    &= \sup_v \inf_{u \in v^\perp}
    \frac{\langle u, A_k u \rangle}{\langle u, u \rangle}
    = \sup_v \inf_{\substack{u \\ \langle G_k^{1/2} u, v \rangle = 0}}
    \frac{\langle u, H_k u \rangle}{\langle u, G_k u \rangle} \\
    &= \sup_v \inf_{u \in v^\perp}
    \frac{\langle u, H_k u \rangle}{\langle u, G_k u \rangle}
    \ge \sup_v \inf_{u \in v^\perp}
    \frac{\langle u, H_0 u \rangle}{\|u\|^2}
    \frac{k^2 \|u\|^2}{\langle u, k^2 G_k u \rangle}
    \ge \tfrac34 k^2,
  \end{align*}
  which for small $k$ is larger than $\zeta_0$.  Here we used
  knowledge of the spectrum of $H_0$: it has zero and $3/4$ as
  isolated eigenvalues and a continuous spectrum $[1,\infty)$.
\end{proof}

\section{The resolvent: a first order system}
\label{system}

In Sections \ref{system}--\ref{keres} we  analyze the
integral kernel of the resolvent of the
operator $\zeta - D_k H_k$, where $\zeta$ is a complex number
 outside the spectrum of $D_k H_k$; see \eqref{reso}.
We start by writing $(\zeta -
D_k H_k) u = f$ as a system of ordinary linear
differential equations
\begin{equation}
\label{eq:geninhom}
  \fvec{u}'(x) = \fvec{A}(\zeta, k , x) \fvec{u}(x) + \fvec{b}(x),
\end{equation}
where
\begin{align}
  \fvec{u} &= (u, u', u'', u''')^T, & \fvec{b} &= (0, 0, 0, -f)^T
 \label{eq:CHyb}
\end{align}
and (for $V$, see \eqref{ell1} )
\begin{equation} \label{eq:CHA}
  \fvec{A} = 
  \begin{pmatrix}
    0 & 1 & 0 & 0 \\
    0 & 0 & 1 & 0 \\
    0 & 0 & 0 & 1 \\
    \zeta - k^2 - k^4 - k^2 V + V'' & 2 V'
    & 1 + 2 k^2 + V & 0
  \end{pmatrix}.
\end{equation}

We define the $x$-independent matrices
 $\fvec{A}_\infty (\zeta ,k ) := \lim_{x \to \pm \infty}
 \fvec{A}( \zeta, k, x)$,
and $\fvec{R}( \zeta ,k, x) := \fvec{A}(\zeta, k , x) -
\fvec{A}_\infty( \zeta ,k)$; the latter
can be bounded by $C e^{-|x|}$
for $(\zeta , k)$ in any compact set.
The positive constant $C$ may depend on the set.

The eigenvalues of $\fvec{A}_\infty$ are the zeros of the polynomial
\[
\zeta - (-\mu^2 + k^2)^2 - (-\mu^2 + k^2)
\]
which are
\begin{equation} \label{eq:mu}
  \mu_j = \pm \sqrt{\hf + k^2 \pm \hf \sqrt{1 + 4 \zeta}}.
\end{equation}
Unfortunately they have somewhat poor analyticity for $k$ and
$\zeta$ near zero.  This problem can be overcome by writing
\begin{equation} \label{eq:zetaoftau}
  \zeta = k^2 + k^4 + (1 + 2 k^2)^2 \tau^2
\end{equation}
and using $\lambda = (k,\tau)$ as the parameters.
Then \eqref{eq:CHA}
and \eqref{eq:mu} become
\[
\fvec{A} = 
\begin{pmatrix}
  0 & 1 & 0 & 0 \\
  0 & 0 & 1 & 0 \\
  0 & 0 & 0 & 1 \\
  (1 + 2 k^2)^2 \tau^2 - k^2 V + V'' & 2 V'
  & 1 + 2 k^2 + V & 0
\end{pmatrix}
\]
and
\begin{equation} \label{eq:muoftau}
  \mu_j = \pm \sqrt{1 + 2 k^2} \sqrt{\tfrac12 \pm
  \tfrac12 \sqrt{1 + 4 \tau^2}}.
\end{equation}
To fix the branches when the second sign is negative, set
\begin{equation} \label{eq:tau-branch}
  \pm\sqrt{\hf - \hf\sqrt{1 + 4 \tau^2}}
  = \frac{\pm \ii \tau}{\sqrt{\hf + \hf\sqrt{1 + 4 \tau^2}}}
\end{equation}  
and choose $\mu_2$ to have the positive sign with the principal branch
in the denominator.  Now the $\mu_i$ are analytic functions of $k$ and
$\tau$ when $k \in \bbC \setminus \pm \ii [1/\sqrt 2, \infty)$ and $\tau
\in \bbC \setminus \pm \ii [\hf,\infty)$.  We are mostly interested in
real $k$.  Then the cuts in the $\tau$ plane correspond to $\zeta \in
(-\infty, -1/4]$, values which we do not need for integrating around
the spectrum of $D_k H_k$.  $\tau \in \pm \ii [0, \hf k] \cup \bbR$ is
mapped onto $[3 k^2/4 - k^6, \infty)$ which by
Lemma~\ref{lm:coarse-spectrum} contains all of the spectrum of $D_k
H_k$ except for the lowest eigenvalue, which should be somewhere
around $\tau \approx \ii k$ when $k$ is small.  For
larger $k$ it is more useful to use the $k^4$ bound: $\tau \in \pm\ii
[0,k/(1+2k^2)] \cup \bbR$ is mapped onto $[k^4,\infty)$ which contains
the entire spectrum.

The number $\tau$ is mapped two-to-one on $\zeta$ and we will mostly
keep $\tau$
in the upper half plane.  There $\Re \mu_2 < 0$ when $k$ is real.
However, we need some results to be valid also in a small complex
neighborhood of the origin for both $k$ and $\tau$, because we will
develop a power series there later.

The number $\mu_1$ will be the eigenvalue with the most
negative real part, $\mu_4 :=
-\mu_1$ and $\mu_3 := -\mu_2$. We have
\begin{equation} \label{eq:murel}
  \begin{gathered}
    \mu_1^2 +\mu_2^2 = 1 + 2 k^2,
    \qquad
    \mu_1\mu_2 = -\ii \tau (1 + 2 k^2),\\
    \mu_1^2 -\mu_2^2 = (1 + 2 k^2) \sqrt{1 + 4 \tau^2} = \sqrt{1 + 4\zeta}.
  \end{gathered}
\end{equation}

We denote by $\fvec{v}_j$ and $\fvec{w}_j$ the right and left eigenvectors
of $\fvec{A}_\infty$: $\fvec{A}_\infty \fvec{v}_j = \mu _j \fvec{v}_j $ and
$\fvec{w}_j \fvec{A}_\infty  = \mu _j \fvec{w}_j $.
The vectors $\fvec{v}_j$ and $\fvec{w}_j$ are easy to express
in terms of $\mu_j$
and are thus also analytic.  Normalize the eigenvectors so that
$(\fvec{v}_j)_1 = (\fvec{w}_j)_4 = 1$.  This results in
\begin{align*}
  \fvec{v}_j &= \begin{pmatrix}
    1, & \mu_j, & \mu_j^2, & \mu_j^3
  \end{pmatrix}^T, \\
  \fvec{w}_j &= \begin{pmatrix}
    \mu_j (\mu_j^2 - 1 - 2 k^2), & \mu_j^2 - 1 - 2 k^2, & \mu_j, & 1
  \end{pmatrix}.
\end{align*}

\section{The resolvent: solutions of the homogeneous equations}

The resolvent for \eqref{eq:geninhom} will be constructed using
the solutions of the corresponding homogeneous equation
and its transposed equation ($\fvec{A}(x):= \fvec{A}(\zeta, k,x)$
and so on):
\begin{align}
  \fvec{u}'(x) &= \fvec{A}(x) \fvec{u}(x) \label{eq:genhom} \\
  \fvec{z}'(x) &= -\fvec{z} (x) \fvec{A}(x). \label{eq:gentransp}
\end{align}
So, this section contains a study of the solutions of
\eqref{eq:genhom} and \eqref{eq:gentransp}.
The motivation for \eqref{eq:gentransp}
is that if $\fvec{u}$ and $\fvec{z}$ are
solutions of \eqref{eq:genhom} and \eqref{eq:gentransp} the product
$\fvec{z}(x) \fvec{u}(x)$ is independent of $x$.

As the coefficient matrix $\fvec{A}(x)$ tends to the
constant $\fvec{A}_\infty$ when
$x \to \infty$, one would also expect the solutions of
\eqref{eq:genhom} and \eqref{eq:gentransp} to tend to the solutions of
the corresponding constant coefficient equations.  Indeed, Pego and
Weinstein show in \cite{PegoWeinstein} that if there is a simple
eigenvalue $\mu_1$ with a smaller real part than the other eigenvalues,
\eqref{eq:genhom} admits a unique solution $\fvec{u}_1^+$ with the property
that $e^{-\mu_1 x} \fvec{u}_1^+(x) \to \fvec{v}_1$ as $x \to \infty$.
Similarly
\eqref{eq:gentransp} has an unique solution $\fvec{z}_1^-$ with $e^{\mu_1 x}
\fvec{z}_1^- (x) \to \fvec{w}_1$ as $x \to -\infty$.

We now examine solutions corresponding to the remaining eigenvalues
$\mu_j$ for $j \ge 2$.  We prove things for $\fvec{u}^+_j$ but the
$\fvec{z}^-_j$ case is similar.  Fix a set of values $\lambda := (k ,
\tau) \subset \bbC^2$ to work with.  (As explained in the previous
section, the dependence of $\fvec{A}$ or $\mu_j$ on $\lambda$ or $k$ or
$\tau$ is analytic in the natural way, if $\lambda$ is restricted to a
suitable domain.)  Choose some $\lambda$-dependent $\mu$ and
substitute $\fvec{u}(x) =: e^{\mu x} \fvec{v} (x)$ in \eqref{eq:genhom}
to get
\begin{equation} \label{eq:bhom}
  \fvec{v}'( x) = (\fvec{B} + \fvec{R}( x)) \fvec{v}( x),
\end{equation}
where $\fvec{B} := \fvec{A}_\infty -\mu$ and $\fvec{R}( x) :=
\fvec{A}( x) - \fvec{A}_\infty$.

We want to divide the eigenvalues of $\fvec{A}_\infty$ into two sets: those
with smaller real part than $\mu$ and the rest.  However, we need some
margins in order to keep things uniform in $\lambda$, because the
eigenvalues may not maintain the same order for all $\lambda$
(see Section~\ref{system}).

Let $E_i := \bigcup_{n=1}^\infty \ker(\mu_i - \fvec{A}_\infty)^n$.
Choose constants $\alpha$ and $\beta$ and a set $I \subset \{ 1,2,3,4
\}$ so that $\Re(\mu_i -\mu) \le \alpha < 0$ when $i
\in I$ and $\Re(\mu_i -\mu) \ge \beta > -\rho$ whenever $i \not\in I$.
The numbers $\alpha$ and $\beta$ are assumed independent of $\lambda$.
Let $\fvec{P}$ be the projection onto $\bigoplus_{i \in I} E_i$ which
commutes with $\fvec{A}_\infty$ and let $\fvec{Q} := \fvec{1} - \fvec{P}$.
Fix some positive $x_0$ and define
\begin{equation} \label{eq:calF}
  \begin{aligned}
    (\cF_P \fvec{v})(x) &:= \int_{x_0}^x e^{(x - {y}) \fvec{B}}
    \fvec{P} \fvec{R}({y}) \fvec{v}({y})
    \,\dee{y}, \\
    (\cF_Q \fvec{v})(x) &:= -\int_x^\infty e^{(x - {y}) \fvec{B}}
    \fvec{Q} \fvec{R}({y}) \fvec{v}({y})
    \,\dee{y}, \ \ \ \
    \cF := \cF_P + \cF_Q
  \end{aligned}
\end{equation}
for bounded continuous functions $\fvec{v} : [x_0,\infty] \to \bbC^4$.
The proof of the following facts is straightforward:
\begin{lemma}
  For sufficiently large $x_0$, $\cF$ is a contraction in the norm
  \[
  \|\fvec{v}\|= \sup_{x \in [x_0, \infty)} |\fvec{v}(x)|.
  \]
\end{lemma}
\begin{corollary}
$(\cF \fvec{v})(x) = \Oh(e^{\max(-\rho, \alpha + \epsilon)x})$ for
large $x$ and any $\epsilon > 0$.
\end{corollary}
Hence, $\fvec{v} = \tilde{\fvec{v}} + \cF \fvec{v}$ can be solved
for $\fvec{v}$ given any $\tilde{\fvec{v}}$.
For such a  solution
\begin{align*}
  (\fvec{v} - \tilde{\fvec{v}})' &= \fvec{P}\fvec{R}v + \fvec{B} \cF_P
  \fvec{v} + \fvec{Q}\fvec{R} \fvec{v} + \fvec{B} \cF_Q \fvec{v} 
  =\fvec{B} (\fvec{v} - \tilde{\fvec{v}}) + \fvec{R}v.
\end{align*}
In particular if $\tilde{\fvec{v}}$ is a bounded solution of
$\tilde{\fvec{v}}' = \fvec{B}
\tilde{\fvec{v}}$ (a constant coefficient equation)
for large $x$ then $\fvec{v}$
will be a solution of \eqref{eq:bhom} with the same asymptotic
behavior as $\tilde{\fvec{v}}$ in the sense that $\fvec{v} -
\tilde{\fvec{v}} = \cF \fvec{v}$ tends
to zero exponentially fast as $x \to \infty$.
\begin{corollary} \label{cor:homsol}
  For each eigenvalue $\mu_i$ there is a solution of \eqref{eq:genhom}
  which behaves asymptotically (as $x \to \infty$) like $e^{\mu_i x}
  \fvec{v}_i$.
\end{corollary}
\begin{proof}
  Pick $\mu =\mu_i$, $\alpha = -\frac58$,
  $\beta = -\tfrac78$,
  $\epsilon = \frac1{16}$ and $\tilde{\fvec{v}}(x) = \fvec{v}_i$.
  It is clear
  that a suitable $I$ can be found and $\cF$ can be used to get a
  solution for $x > x_0$, which can then be extended to the whole real
  line.
\end{proof}

In general the solutions of Corollary~\ref{cor:homsol} are not unique
even after fixing normalization of $\fvec{v}_i$ because one can add similar
solutions corresponding to $\mu_j$ with smaller real parts than $\mu_i$.

\begin{corollary} \label{cor:homsolanalytic}
  If the $\mu_i$ and $\fvec{v}_i$ are analytic functions of $\lambda$ in some
  domain of $\bbC^2$ and we can fix the $I$ in the preceding
  proof uniformly for
  all $\lambda$, then the solutions in Corollary~\ref{cor:homsol} are
  analytic in this domain when evaluated at some fixed $x$.  If $\fvec{u}_i$
  is such a solution then $|e^{-\mu_i(\lambda) x} \fvec{u}_i(\lambda;x) -
  \fvec{v}_i(\lambda)| < C e^{-\frac\rho2 x}$ for $x > 0$ and $\lambda$ in
  compact subsets of the domain ($C$ depends on the
  subset).
\end{corollary}

By Corollary~\ref{cor:homsol}, for $j \in \{1, 2, 4\}$ there
is a $\fvec{u}^+_j$ which solves the homogeneous
equation $\partial_x \fvec{u}^+_j = \fvec{A}
\fvec{u}^+_j$ and behaves like $e^{\mu_j x} \fvec{v}_j$ as $x \to \infty$.
There is also a $\fvec{z}^-_j$ which solves the transposed equation
$\partial_x \fvec{z}^-_j =
-\fvec{z}^-_j \fvec{A}$ and behaves like $e^{-\mu_j x} \fvec{w}_j $
as $x \to -\infty$.  It
would also be possible to define $\fvec{u}^+_3$
and $\fvec{z}^-_3$ in this way but
that would not be very useful because $\fvec{A}_\infty$ is defective at
$\tau = 0$: the eigenvectors $\fvec{v}_2$ and $\fvec{v}_3$ collide and we would
not have a set of four linearly independent solutions.  Thus with some
abuse of notation we require instead
\[
\fvec{u}^+_3( x) \sim \frac{e^{\mu_3 x} \fvec{v}_3 - e^{\mu_2 x}
\fvec{v}_2}{\mu_3 -\mu_2}
\quad \text{and} \quad
\fvec{z}^-_3( x) \sim \frac{e^{-\mu_3 x} \fvec{w}_3 - e^{-\mu_2 x}
\fvec{w}_2}{\mu_3 -\mu_2}
\]
as $x \to \infty$ and $x \to -\infty$, respectively.  As $\mu_3 -\mu_2
\to 0$ these converge (pointwise in $x$) to solutions with linear
asymptotes.

Our expression for the integral kernel of $(\zeta - D_k H_k)^{-1}$
will contain only $\fvec{u}^+_j$ and $\fvec{z}^-_j$ for $j \in \{1, 2\}$.
However,
the other two values of $j$ are needed for understanding the behavior
of the kernel.

We summarize the properties of the solutions of the homogeneous
equations in the following theorem.
\begin{theorem} \label{th:yz}
  \begin{equation} \label{eq:mucond}
    \Re(\mu_1 \pm \mu_2) < -\tfrac58
    \quad\text{and}\quad
    \Re(\mu_2) < \tfrac1{32}.
  \end{equation}
  Then \eqref{eq:genhom} with \eqref{eq:CHA} has solutions $\fvec{u}^+_i$
  such that $|\fvec{u}^+_i(x) - e^{\mu_i x} \fvec{v}_i| < C |
  e^{(\mu_i - \hf) x}|$   for $i \in   \{1,2,4\}$ and
  \[
  |\fvec{u}^+_3(x) - \frac{e^{\mu_3 x} \fvec{v}_3 - e^{\mu_2 x}
  \fvec{v}_2}{2\mu_3}|   < C |e^{(\mu_3 - \hf) x}|
  \]
  when $x$ is bounded from below.  Furthermore,
  \begin{align*}
    |\partial_x^n \fvec{u}^+_i(x) -\mu_i^n e^{\mu_i x} \fvec{v}_i|
    &< C |e^{(\mu_i - \hf) x}|,
    \qquad n \in \{1,2\},\quad i \ne 3 \\
    |\partial_x \fvec{u}^+_3(x) - \hf (e^{\mu_3 x} \fvec{v}_3 +
    e^{\mu_2 x} \fvec{v}_2)|
    &< C |e^{(\mu_3 - \hf) x}|, \\
    |\partial_x^2 \fvec{u}^+_3(x) - \hf (\mu_3 e^{\mu_3 x}
    \fvec{v}_3 +\mu_2 e^{\mu_2 x} \fvec{v}_2)|
    &< C |e^{(\mu_3 - \hf) x}|.
  \end{align*}
  Similarly, \eqref{eq:gentransp} has solutions $\fvec{z}^-_i$ such that
  $|\fvec{z}^-_i(x) - e^{-\mu_i x} \fvec{w}_i| < C e^{(\hf -\mu_i) x}$
  for $i \in   \{1,2,4\}$ and
  \[
  |\fvec{z}^-_3(x) - \frac{e^{-\mu_3 x} \fvec{w}_3 - e^{-\mu_2 x}
  \fvec{w}_2}{2\mu_3}|   < C |e^{(\hf -\mu_3) x}|
  \]
  when $x$ is bounded from above.  The corresponding estimates for the
  derivatives hold.
  
  For $i \in \{1,2,4\}$ the functions $\fvec{u}^+_i(x)$ and $\fvec{z}^-_i(x)$
  are analytic in
  $\lambda$ wherever the assumptions hold. The functions
  $\fvec{u}^+_3(x)$ and $\fvec{z}^-_3(x)$
  are continuous and when
  \begin{equation} \label{eq:ancond}
    \Re(\mu_2) > -\tfrac14
  \end{equation}
  also analytic. The constant $C$ above depends on $\lambda$
  but can be fixed in
  any compact subset.
\end{theorem}
The proof is basically just application of Corollaries
\ref{cor:homsol} and \ref{cor:homsolanalytic} with some additional
details for the $i = 3$ case.  There we need to use different choices
for the set $I$ depending whether $\Re(\mu_2)$ is near zero or not.
The results can be glued together with a partition of the unity but
analyticity is then lost.  See \cite{timo} for more details.

Specifying the assumptions in terms of $\mu_1$ and $\mu_2$ was perhaps
a bit opaque.  However, there are two cases which interest us.  The
first is $\lambda \approx 0$.  In a small enough neighborhood of the
origin both \eqref{eq:mucond} and \eqref{eq:ancond} clearly hold.  The
second case is when $k$ is real and $\tau$ in the upper half plane.
Then $\mu_2$ is in the left half plane.  For the first inequality of
\eqref{eq:mucond} use \eqref{eq:murel} to get $(\mu_1 \pm \mu_2)^2 =
(1 + 2 k^2) (1 \mp 2 \ii \tau)$.  From this we see that the inequality
is equivalent to $\tau$ lying beneath an upwards opening parabola with
apogee at $\ii (128 k^2 + 39)/(256 k^2 + 128)$.  This is always above
the values corresponding to the spectrum.

To end this section we still note the following fact, which is
a consequence of the symmetry of the Cahn--Hilliard equation
under the reflection $x \mapsto -x$:
\begin{lemma} \label{lm:yzopposite}
  Under the assumptions \eqref{eq:mucond} there are also
  solutions $\fvec{u}^-_j$ and $\fvec{z}^+_j$ of \eqref{eq:genhom}
  and \eqref{eq:gentransp}, respectively, with
  prescribed behavior for $x \to -\infty$ and $x \to \infty$: (e.g.
  $\fvec{u}^-_1(x) \sim e^{-\mu_1 x} \fvec{v}_4$ as $x \to -\infty$).
  These can be expanded as linear combinations of $\fvec{u}^+_j$
  and $\fvec{z}^-_j$.  The coefficients of this expansion are continuous
  and, when \eqref{eq:ancond} holds, also analytic.
\end{lemma}

\section{The resolvent: a resolvent formula}
\label{r-formula}

We now want to write solutions of \eqref{eq:geninhom},
that is, of $\fvec{u}' = \fvec{A} \fvec{u} + \fvec{b}$,
using the solutions of the
homogeneous equations. We will use the following resolvent
formula:
\begin{equation} \label{eq:vecres}
  \fvec{u}(x) = \int_{-\infty}^x \fvec{u}^+(x)
  (\Omega^+)^{-1} \fvec{z}^-({y}) \fvec{b}({y}) \,\dee{y}
  + \int_\infty^x \fvec{u}^-(x) (\Omega^-)^{-1} \fvec{z}^+({y})
  \fvec{b}({y}) \,\dee{y},
\end{equation}
where $\fvec{u}^\pm = (\fvec{u}^\pm_1, \fvec{u}^\pm_2)$, $\fvec{z}^\mp = \left(
  \begin{smallmatrix}
    \fvec{z}_1^\mp \\ \fvec{z}_2^\mp
  \end{smallmatrix}
\right)$ and $\Omega^\pm = \fvec{z}^\mp \fvec{u}^\pm$.
These products are independent
of $x$ and it is also easy to see that $\Omega^- = -\Omega^+$.  For
simplicity we will from now on denote $\Omega^+$ just by $\Omega$.
For \eqref{eq:vecres} to make sense $\Omega$ of course needs to be
invertible.
\begin{theorem} \label{th:Omegainv}
  Assume $\Re\mu_1 < \Re\mu_2 < 0$ and $\zeta \in \rho(D_k H_k)$ (the
  resolvent set).  Then $\Omega = \fvec{z}^- \fvec{u}^+$ is invertible.
\end{theorem}
\begin{proof}
  Assume the contrary: let $\Omega (\alpha_1, \alpha_2)^T = 0$, i.e.,
  $\fvec{z}^-_i \fvec{u} = 0$ for $i \in \{1,2\}$,
  $\fvec{u} := \alpha_1 \fvec{u}^+_1 + \alpha_2
  \fvec{u}^+_2 \ne 0$. The function $\fvec{u}$ can also be written as
  $\sum_{j=1}^4 \beta_j
  \fvec{u}^-_j$.  We have
  \begin{align*}
    e^{\mu_i x} \fvec{z}^-_i(x) &\to \fvec{w}_i, & e^{-\mu_i x}
    \fvec{u}^-_{5-i}(x) &\to    \begin{cases}
      \frac1{2\mu_3} \fvec{v}_2 & \text{when $i = 2$,} \\
      \fvec{v}_i & \text{otherwise}
    \end{cases}
  \end{align*}
  as $x \to -\infty$.  Thus $\fvec{z}^-_1(x) \fvec{u}^-_j(x) \to
  \delta_{4j} \fvec{w}_1
  \fvec{v}_1$ but the left side is actually independent of $x$.
  Since $\mu_1$
  is a simple eigenvalue we must have $\fvec{w}_1 \fvec{v}_1 \ne 0$,
  consequently
  $\fvec{z}^-_1 \fvec{u} = 0$ implies $\beta_4 = 0$.
  In a similar vein $\fvec{z}^-_2 \fvec{u} =
  0$ implies $\beta_3 = 0$.  Hence $\fvec{u} =
  \beta_1 \fvec{u}^-_1 + \beta_2 \fvec{u}^-_2$
  decreases exponentially at $\pm\infty$, giving a nontrivial $L^2$
  solution to $(\zeta - D_k H_k) u = 0$.  Such a solution
  was assumed not to exist.
\end{proof}

We omit the proof that \eqref{eq:vecres} is actually a solution
of \eqref{eq:geninhom}.

The solutions $\fvec{u}^\pm_j$ can be written in terms of a
scalar valued function $U^\pm_j$.  
A straightforward computation yields:
\begin{lemma}
  \begin{align*}
    \fvec{u}^\pm_j &=
    \begin{pmatrix}
      U^\pm_j, & \partial_x U^\pm_j,
      & \partial_x^2 U^\pm_j, & \partial_x^3 U^\pm_j
    \end{pmatrix}^T \\
    \fvec{z}^\mp_j &= 
    \begin{pmatrix}
      (H_k \partial_x + k^2 \partial_x - V') Z^\mp_j,
      & -(H_k + k^2) Z^\mp_j,
      & -\partial_x Z^\mp_j, & Z^\mp_j
    \end{pmatrix}
  \end{align*}
with $(\zeta - D_k H_k) U^\pm_j = 0$ and $(\zeta - H_k D_k) Z^\mp_j =
0$.  For $j \in \{1,2,4\}$ we have $\lim_{x \to \infty} e^{-\mu_j x}
U^+_j(x) = \lim_{x \to -\infty} e^{\mu_j x} Z^-_j(x) = 1$
while     $\lim_{x
\to \infty} e^{-\mu_3 x} U^+_3(x) - \frac{e^{2\mu_2 x} - 1}{2\mu_2} =
\lim_{x \to -\infty} e^{\mu_3 x} Z^-_3(x) - \frac{e^{-2\mu_2 x} - 1}{2
\mu_2} = 0$ (when $\mu_2 = 0$, replace the fractions by their limits,
i.e., $x$ and $-x$).
\end{lemma}

Recalling \eqref{eq:CHyb} we get for the original equation:
\begin{theorem} \label{th:reskernel}
  Under the assumptions of Theorem~\ref{th:Omegainv} the integral
  kernel of the resolvent is given by
  \begin{equation} \label{eq:reskernel}
    R(x,{y}) := (\zeta - D_k H_k)^{-1}(x,{y}) = 
    \begin{cases}
      -U^+(x) \Omega^{-1} Z^-({y}) & \text{for ${y} < x$,} \\
      -U^+(-x) \Omega^{-1} Z^-(-{y}) & \text{for ${y} > x$,}
    \end{cases}
  \end{equation}
  where $U^+ = (U^+_1, U^+_2)$ and $Z^- = (Z^-_1, Z^-_2)^T$.
\end{theorem}
Note that from $\fvec{u}^+( x) \Omega^{-1} \fvec{z}^-(x) - \fvec{u}^-( x)
\Omega^{-1} \fvec{z}^+(
x) = 1$ it follows in particular that the $(i,j)$ component of the
left hand side is $0$ for $i < j$.  From this we see that our
resolvent kernel has continuous derivatives with respect to $x$ or
${y}$ up to total order two.

\section{Estimates for the resolvent}
\label{keres}

The leading terms of the semigroup will arise from the resolvent with
small $k$ and $|\zeta|$ values. Deriving estimates for the resolvent
kernel \eqref{eq:reskernel} for these parameter values is the main
task of this section, leading to Lemma~\ref{lm:Omega} and
Theorem~\ref{th:F}.  After this we consider large parameter values
briefly in Theorem~\ref{th:respert}.

We use \eqref{eq:zetaoftau} and assume in the following that $k <
\epsilon$, $|\tau| < \epsilon$ for a small enough $\epsilon$.  In
particular we assume that the smallest eigenvalue is isolated (see
Lemma~\ref{lm:coarse-spectrum}) and that the analyticity
condition~\eqref{eq:ancond} and the other assumptions of
Theorem~\ref{th:yz} are satisfied.  We will proceed to develop
$\Omega$ of \eqref{eq:reskernel} into a power series to get some
explicit leading terms for the resolvent.  Thus this section is mostly
about computing derivatives of things at $k = \tau = 0$.

We denote the solutions of the homogeneous equation
at $k = \tau = 0$ with a
r$\zeroth{\text\i}$ng.  $\zeroth \mu_1 = -1$, $\zeroth \mu_2 = 0$ and
\begin{equation} \label{eq:yz}
  \begin{aligned}
    \zeroth U_1^+(x) &= \frac1{4 \cosh(\frac x2)^2} \\
    \zeroth U_2^+(x) &= \frac{-1 - 6 e^x + 5 e^{2 x} +
      2 e^{3 x} + 6 e^{2 x} x}{2 e^x (1 + e^x)^2} \\
    \zeroth Z_1^-(x) &= \log(e^x + 1) \\
    \zeroth Z_2^-(x) &= 1.
  \end{aligned}
\end{equation}
By a mechanical computation we get
\[
\zeroth \Omega = 
\begin{pmatrix}
  0 & 1 \\
  0 & 0
\end{pmatrix}.
\]

The solutions which are asymptotically equal to $x \mapsto |x|$, i.e.,
$\zeroth U^+_3(x)/x \to 1$ as $x \to \infty$ and $\zeroth
Z^-_3(x)/x \to -1$ as $x \to -\infty$ can also be computed: $\zeroth
Z^-_3(x) = -x$ but the expression for $\zeroth U^+_3$ is lengthy and
we omit it.  The following products will be needed later:
\begin{equation} \label{eq:zy3}
  \begin{pmatrix}
    \zeroth{\fvec{z}}^-_1 \\ \zeroth{\fvec{z}}^-_2 \\ \zeroth{\fvec{z}}^-_3
  \end{pmatrix}
  \zeroth{\fvec{u}}^+_3 =
  \begin{pmatrix}
    0 \\ -1 \\ 0
  \end{pmatrix}
  \qquad \text{and} \qquad
  \zeroth{\fvec{z}}^-_3
  \begin{pmatrix}
    \zeroth{\fvec{u}}^+_1 & \zeroth{\fvec{u}}^+_2 & \zeroth{\fvec{u}}^+_3
  \end{pmatrix}
  = \begin{pmatrix}
    0 & -1 & 0
  \end{pmatrix}.
\end{equation}
We will also also need the rapidly growing solution
$
\zeroth U^+_4(x) = 4 \cosh(\frac x2)^2.
$

\paragraph*{First order:}
Define $\fvec{v}^+_j(x) = e^{-\mu_j x} \fvec{u}^+_j(x)$ and
$\fvec{w}^-_j(x) = e^{\mu_j x}
\fvec{z}^-_j(x)$ for $j \in \{1, 2\}$.  By Corollary~\ref{cor:homsolanalytic}
they are analytic functions of $\lambda$ for each fixed $x$ and
satisfy $|\fvec{v}^+_j(x) - \fvec{v}_j| < C e^{-x/2}$ for $x > 0$ and
$|\fvec{w}^-_j(x) -
\fvec{w}_j| < C e^{x/2}$ for $x < 0$ where $C$ can be fixed independently
of $k$ and $\tau$.  They will also satisfy the obvious differential
equations, which we now differentiate, then set $k = \tau = 0$:
\begin{align*}
  \partial_x \partial_\lambda \fvec{v}^+_j &=
  (\fvec{A} - \mu_j) \partial_\lambda \fvec{v}^+_j
  - \partial_\lambda \mu_j \fvec{v}^+_j \\
  \partial_x \partial_\lambda \fvec{w}^-_j
  &= - \partial_\lambda \fvec{w}^-_j (\fvec{A} - \mu_j)
  + \partial_\lambda \mu_j \fvec{w}^-_j
\end{align*}
(note that $\partial_\lambda \fvec{A} = 0$).  These are just inhomogeneous
versions of the equations satisfied by $\fvec{v}^+_j$ and $\fvec{w}^-_j$.  The
``initial conditions'' are again asymptotic:
$|\partial_\lambda \fvec{v}^+_j(x) -
\partial_\lambda \fvec{v}_j| < C e^{-x/2}$ and
 $|\partial_\lambda \fvec{w}^-_j(x) - \partial_\lambda \fvec{w}_j|
< C e^{x/2}$ by Cauchy's estimates.
 The first component of $\partial_\lambda
\fvec{v}_j$ and the last component of $\partial_\lambda \fvec{w}_j$ are
zero because of the
normalization used.

Using all available information developed until now, we
end up with these reasonably simple expressions:
\begin{align}
  \partial_\lambda \fvec{u}^+_j &= \partial_\lambda
  \mu_j \zeroth{\fvec{u}}^+_3,
  & \partial_\lambda \fvec{z}^-_j &= \partial_\lambda
  \mu_j \zeroth{\fvec{z}}^-_3 \label{eq:dyz}
\end{align}
and, using \eqref{eq:zy3},
\[
\partial_\lambda \Omega = \partial_\lambda \fvec{z}^-
\fvec{u}^+ + \fvec{z}^- \partial_\lambda \fvec{u}^+ =
\begin{pmatrix}
  0 & 0 \\ 0 & -2 \partial_\lambda \mu_2
\end{pmatrix},
\]
with $\partial_\lambda \mu_1 = \partial_k \mu_2 = 0$ and
$\partial_\tau \mu_2 = \ii$.

\paragraph*{Second order:}
Above we obtained non-zero leading terms for the right
column of $\Omega$.
We go on with the left column in order to eventually compute some non-zero
terms for $\det\Omega$.
We have to deal with
\[
\partial_\lambda^2 (\fvec{z}^-_i \fvec{u}^+_1)
= \partial_\lambda^2 \fvec{z}^-_i \fvec{u}^+_1 + \fvec{z}^-_i
\partial_\lambda^2 \fvec{u}^+_1,
\]
since $\partial_\lambda \fvec{u}^+_1 = 0$ at $\lambda = 0$.  It is
convenient to estimate the terms on the right hand side at separate
values of $x$, which is possible with the following trick:
\[
\partial_x (\partial_\lambda^2 \fvec{z}^-_i \fvec{u}^+_1)
= -\partial_\lambda^2 (z^-_i \fvec{A}) \fvec{u}^+_1 +
\partial_\lambda^2 \fvec{z}^-_i \fvec{A} \fvec{u}^+_1
= -z^-_i \partial_\lambda^2 \fvec{A} \fvec{u}^+_1
\]
(recalling $\partial_\lambda \fvec{A} = 0$).  Now we can write
\[
\partial_\lambda^2 (\fvec{z}^-_i \fvec{u}^+_1)
= (\partial_\lambda^2 \fvec{z}^-_i \fvec{u}^+_1)(x_0) +
(\fvec{z}^-_i \partial_\lambda^2 \fvec{u}^+_1)(x_1)
- \int_{x_0}^{x_1} \fvec{z}^-_i \partial_\lambda^2 \fvec{A} \fvec{u}^+_1 \,\dee x.
\]
Taking the limits $x_0 \to -\infty$ and $x_1 \to \infty$
and using the asymptotic behavior of 
$\zeroth{\fvec{u}}^+_1$ and other functions involved everything
except the integral vanishes and we get
\begin{align*}
  \partial_k^2 (\fvec{z}^- \fvec{u}^+_1) &=
  \begin{pmatrix}
    -\tfrac73 \\ -2
  \end{pmatrix}, &
  \partial_k \partial_\tau (\fvec{z}^- \fvec{u}^+_1) &= 0, &
  \partial_\tau^2(\fvec{z}^- \fvec{u}^+_1) &=
  \begin{pmatrix}
    -2 \\ -2
  \end{pmatrix}.
\end{align*}

\paragraph*{Third order:}
Similarly we obtain $\partial_\lambda^3 (\fvec{z}^-_2 \fvec{u}^+_1) = 0$.

Collecting everything from above we get
\begin{lemma} \label{lm:Omega}
  \begin{align}
    \Omega &=
    \begin{pmatrix}
      -\frac76 k^2 - \tau^2  & 1 \\
      -k^2 - \tau^2 & -2 \ii \tau
    \end{pmatrix}
    +
    \begin{pmatrix}
      \Oh(\lambda^3) & \Oh(\lambda^2) \\
      \Oh(\lambda^4) & \Oh(\lambda^2)
    \end{pmatrix}, \notag \\
    \det\Omega &=
    k^2 + \tau^2 + 2 \ii \tau^3 + \tfrac73 \ii \tau k^2 + \Oh(k^4) + \Oh(\tau^4).
    \label{eq:detOmega}
  \end{align}
\end{lemma}
According to \eqref{eq:reskernel}, $R(x,{y}) = R(-x,-{y})$.  Thus it
suffices to consider the case ${y} < x$.  Then
\[
R( x, {y}) = \frac{-1}{\det\Omega} U^+( x) \mho Z^-( {y}).
\]
where
\[
\mho =
\begin{pmatrix}
  -2 \ii \tau  & -1 \\
  k^2 + \tau^2 & -\frac76 k^2 - \tau^2
\end{pmatrix}
+
\begin{pmatrix}
  \Oh(\lambda^2) & \Oh(\lambda^2) \\
  \Oh(\lambda^4) & \Oh(\lambda^3)
\end{pmatrix}.
\]
In our small neighborhood $(k,\tau) \approx 0$ the only possible
singularity is that $\det \Omega$ may become zero, producing a pole in
the resolvent.  To examine the $x$ and ${y}$ dependence of the
resolvent it is easier to work with $F(\lambda; x,{y}) := - U^+( x)
\mho Z^-( {y})$ which has no such singularity.
\begin{multline*}
  F( \lambda; x,y) = U^+_1( \lambda; x)
  (Z^-_2( \lambda; y) + 2 \ii \tau Z^-_1( \lambda; y))\\
  + \sum_{i=1}^2 \sum_{j=1}^2
  \gamma_{ij}(\lambda) U^+_i( \lambda; x) Z^-_j( \lambda; y),
\end{multline*}
where the $\gamma_{ij}$ are analytic functions of $\lambda = (k,\tau)$
and $\Oh(\lambda^2)$.  It turns out that we need to look at the
$U^+_2 Z^-_1$ term a bit more carefully so let us write it more
explicitly:
\begin{multline} \label{eq:U+Z-}
  F = U^+_1 Z^-_2 + 2 \ii \tau U^+_1 Z^-_1
  - (k^2 + \tau^2 + \Oh(\lambda^4)) U^+_2 Z^-_1 \\
  + \sum_{j=1}^2 \gamma_{1j} U^+_1 Z^-_j + \gamma_{22} U^+_2 Z^-_2.
\end{multline}
As these expressions for $F$ get longer we drop the arguments to
reduce clutter.  $U$ is always evaluated at $x$, $Z$ at $y$ and
everything depends on $\lambda$.  Also, $U^-_i(x) := U^+_i(-x)$ and
$Z^+(y) := Z^-(-y)$.

When $y < 0 < x$ it is easy to bound \eqref{eq:U+Z-} by using
Theorem~\ref{th:yz} but in the other cases where $y < x$ we do not
know much about the behavior of either $U^+_i(x)$ or $Z^-_j(y)$.  We
can get around this by using Lemma~\ref{lm:yzopposite}.

When $0 < y < x$ write
\[
Z^-_i(\lambda; y) = \sum_{j=1}^4 b_{ij}(\lambda) Z^+_j(\lambda; y).
\]
Explicit computation yields
\begin{align*}
  b_{1j} &= \delta_{1j} + \delta_{3j} + \Oh(\lambda) \\
  b_{2j} &= \delta_{2j} - 2 \ii \tau \delta_{3j} + \Oh(\lambda^2).
\end{align*}
Thus
\[
F = U^+_1 Z^+_2 + 2 \ii \tau U^+_1 Z^+_1
+ \sum_{i=1}^2 \sum_{j=1}^4 \beta_{ij} U^+_i Z^+_j
\]
for some coefficients $\beta_{ij}$, which are analytic functions of
$\lambda$ and $\Oh(\lambda^2)$.  $F$ must be a bounded function at least
in the resolvent set.  Hence $\beta_{24}(\lambda)$ must vanish for all
$\lambda$.  The $j=3$ terms, however, are a bit tricky: we have
\[
|e^{-\mu_3 y} Z^+_3(y) - \frac{e^{2 \mu_2 y} - 1}{2 \mu_2}| < C e^{-\hf y}
\]
and $\mu_2 \approx \ii \tau$.  $\beta_{23} = -k^2 - \tau^2 +
\Oh(\lambda^3)$, which is more important than $\beta_{13}$ because of
the rapid decay of $U^+_1$.  Let us write $F$ a bit more explicitly
and also bring in the $U^+_2 Z^+_1$ term:
\begin{multline} \label{eq:U+Z+}
  F = U^+_1 Z^+_2 + 2 \ii \tau U^+_1 Z^+_1
  -(k^2 + \tau^2 + \Oh(\lambda^3)) (U^+_2 Z^+_1 + U^+_2 Z^+_3) \\
  + \sum_{j=1}^4 \beta_{1j} U^+_1 Z^+_j + \beta_{22} U^+_2 Z^+_2.
\end{multline}

When $y < x < 0$ we do the same thing but with $U$:
\begin{align*}
U^+_i &= \sum_{j=1}^4 a_{ij} U^-_j, \\
a_{1j} &= \delta_{1j} + \Oh(\lambda^2), \\
a_{2j} &= 8 \delta_{1j} - \delta_{2j} - \hf \delta_{4j} + \Oh(\lambda).
\end{align*}
Plugging this in we get
\[
F = U^-_1 Z^-_2 + 2 \ii \tau U^-_1 Z^-_1
+ \sum_{i=1}^4 \sum_{j=1}^2 \alpha_{ij} U^-_i Z^-_j
\]
with some coefficients $\alpha_{ij}(\lambda) = \Oh(\lambda^2)$.  Again
$\alpha_{42}$ must vanish to keep the kernel bounded.  To compute the
$U^-_3 Z^-_2$ term coefficient to second order it seems that we would need
$a_{13}$ to second order, which seems difficult to compute.  However,
we can get around this difficulty by using twice continuous
differentiability of $F$ when $y \approx x \ll 0$.  The limits
\begin{align*}
  \lim_{y \to x-} F(x,y)
  &= \alpha_{32} U^-_3(x) Z^-_2(x) + \alpha_{41} U^-_4(x) Z^-_1(x)
  + \Oh(e^{2 \mu_2 |x|}) \\
  \lim_{y \to x+} F(x,y)
  &= \lim_{y \to x+} F(-x,-y) \\
  &= \beta_{23} U^+_2(-x) Z^+_3(-x) + \beta_{14} U^+_1(-x) Z^+_4(-x)
  + \Oh(e^{2 \mu_2 |x|}).
\end{align*}
must be equal.  Working in the region where $\Re \mu_2 < 0$ and taking
the limit $x \to -\infty$ yields
\[
\frac{\alpha_{32}}{2\mu_3} + \alpha_{41}
= \frac{\beta_{23}}{2\mu_3} + \beta_{41}.
\]
Repeat the same for $\p_x^2 F$:
\begin{align*}
  \lim_{y \to x-} \p_x^2 F(x,y)
  &= \alpha_{32} \p_x^2 U^-_3(x) Z^-_2(x)
  + \alpha_{41} \p_x^2 U^-_4(x) Z^-_1(x)
  + \Oh(e^{2 \mu_2 |x|}) \\
  \lim_{y \to x+} \p_x^2 F(x,y)
  &= \lim_{y \to x+} \p_x^2 F(-x,-y) \\
  &= \beta_{23} \p_x^2 U^+_2(-x) Z^+_3(-x)
  + \beta_{14} \p_x^2 U^+_1(-x) Z^+_4(-x) + \Oh(e^{2 \mu_2 |x|}),
\end{align*}
hence
\[
\frac{\mu_3}2 \alpha_{32} + \mu_4^2 \alpha_{41}
= \frac{\mu_3}2 \beta_{23} + \mu_1^2 \beta_{14}.
\]
The two equations for $\alpha_{32}$ and $\alpha_{41}$ are linearly
independent (recall \eqref{eq:murel}) and their obvious solution is
$\alpha_{ij} = \beta_{ji}$.  Thus we get
\begin{multline} \label{eq:U-Z-}
  F = U^-_1 Z^-_2 + 2 \ii \tau U^-_1 Z^-_1
  - (k^2 + \tau^2 + \Oh(\lambda^3)) (U^-_3 Z^-_2 - U^-_2 Z^-_1) \\
  + \sum_{i \in \{1,3,4\}} \alpha_{i1} U^-_i Z^-_1
  + \sum_{i=1}^2 \alpha_{i2} U^-_i Z^-_2.
\end{multline}

To combine the expressions for $F$ in different regions we denote by
$f_{ij}(x,y)$ the part containing $U^\pm_i(x) Z^\pm_j(y)$ in
\eqref{eq:U+Z-}, \eqref{eq:U+Z+} or \eqref{eq:U-Z-}.  We shall also
get rid of $U^-_i$ and $Z^-_j$ by using absolute values.  We write
\begin{equation} \label{eq:Ff}
  F = \sum_{i=1}^4 \sum_{j=1}^4 f_{ij}
\end{equation}
with, e.g., $f_{11}(x,y) = (2 \ii \tau + \Oh(\lambda^2)) U^+_1(|x|)
Z^+_1(|y|)$.

Thus we have written $F$ in terms of $U^+_i$ and $Z^+_i$.  For small
$\lambda$ these functions can be approximated by their explicit forms
at $\lambda=0$, given by \eqref{eq:yz}:
\begin{lemma} \label{lm:UZ}
  \begin{align*}
    U^+_1(x) &= e^{(\mu_1 + 1) x} \zeroth U^+_1(x)
    + \Oh(\lambda^2 e^{(\mu_1 - \hf) x}) \\
    U^+_2(x) &= e^{\mu_2 x} \zeroth U^+_2(x)
    + \Oh(\lambda e^{(\mu_2 - \hf) x}) \\
    Z^+_1(y) &= e^{(\mu_1 + 1) y} \zeroth Z^+_1(y)
    + \Oh(\lambda^2 e^{(\mu_1 - \hf) y}) \\
    Z^+_2(y) &= e^{\mu_2 y} + \Oh(\lambda^2 e^{(\mu_2 - \hf) y}).
  \end{align*}
\end{lemma}
\begin{proof}
  By Theorem~\ref{th:yz}
  \[
  U^+_i(x) = e^{(\mu_i - \zeroth \mu_i) x} \zeroth U^+_i(x) + \Oh(\lambda
  e^{(\mu_i - \hf) x})
  \]
  for $i \in \{1,2\}$ and similarly for the $Z^+_i$.  Recall that
  $\zeroth Z^+_2(x) = 1$.  In some cases the first
  $\lambda$-derivatives of $e^{-\mu_i x} U^+_i(x)$ and $e^{-\mu_i y}
  Z^+_i( y)$ also vanish at $\lambda=0$, hence the remainders are
  $\Oh(\lambda^2)$ rather than $\Oh(\lambda)$.
\end{proof}
By Theorem~\ref{th:yz}
\[
\left| U^+_3(x) - \frac{e^{\mu_2 x} - e^{-\mu_2 x}}{2 \mu_2} \right|
< C e^{(\mu_3 - \hf) x},
\]
similarly for $Z^+_3$.  We also need a couple of derivatives with
respect to $y$:
\begin{lemma} \label{lm:dZ}
  For $n \in \{1,2\}$
  \begin{align*}
    |\p_y^n Z^+_1 - \p_y^n \zeroth Z^+_1|
    &< C |\lambda^2 e^{-\hf y}|, \\
    |\p_y^n Z^+_2 - \mu_2^n e^{\mu_2 y}|
    &< C|\lambda^2 e^{(\mu_2 - \hf) x}|, \\
    |\p_y^n Z^+_3 - \hf(\mu_3^{n-1} e^{\mu_3 y} + \mu_2^{n-1} e^{\mu_2 y})|
    &< C|\lambda e^{(\mu_3 - \hf) x}|.
  \end{align*}
\end{lemma}
\begin{proof}
  $e^{-\mu_2 y} \p_y^n Z^+_2(y) - \mu_2^n$ is known to be
  $\Oh(e^{-x/2})$ by Theorem~\ref{th:yz} and vanishes at $\lambda = 0$.
  Its $\lambda$-derivative at $\lambda = 0$ can be computed (we
  computed $\p_\lambda Z^+_2$ earlier) and also vanishes.  $Z^+_3$ is
  similar except without the $\lambda$-derivative.  For $Z^+_1$ use
  \[
  |e^{-\mu_1 y} \p_y^n Z^+_1 - e^y \p_y^n \zeroth Z^+_1
  - (\mu_1^n - (-1)^n)| < C |\lambda^2 e^{-\hf y}|.
  \]
\end{proof}

To simplify further we use $\mu_2 = -\mu_3 = \ii \tau (1 + \Oh(\lambda^2))$.
Let $c$ be an upper bound for the $\Oh(\lambda^2)$ term, assumed to be
conveniently small.  If $\tau$ is in the sector $\{\Im(\tau) > 2 c
|\tau|\}$ and $y > 0$ we have, denoting $z := \ii \tau y$,
\begin{gather*}
  |e^{\mu_2 y}| < |e^{z + c |z|}| < |e^{\hf \ii \tau y}|, \\
  |e^{\mu_2 y} - e^{\ii \tau y}| = |e^{z} (e^{\Oh(\lambda^2) z} - 1)|
  < C |\lambda^2 e^{z + c |z|}| < C |\lambda^2 e^{\hf \ii \tau y}|.
\end{gather*}
If in addition $x - y > 0$
\[
  \left|
    \frac{e^{\mu_2 (x + y)} - e^{\mu_2 (x - y)}}{2 \mu_2}
    - \frac{e^{\ii \tau (x + y)} - e^{-\ii \tau (x - y)}}{2 \ii \tau}
  \right|
  < C \left| \frac{\lambda^2}\tau e^{\hf \ii \tau (x - y)} \right|.
\]

Using these results we now estimate $\p_y^n f_{ij}$ for the $f_{ij}$
of \eqref{eq:Ff} and $n \in \{0,1,2\}$.  In the case $y < 0 < x$ we
have from \eqref{eq:U+Z-} by Lemmas \ref{lm:UZ} and \ref{lm:dZ} the
following terms and estimates:
\begin{equation} \label{eq:f-common}
  \begin{aligned}
    \p_y^n f_{12} &= \zeroth U^+_1(x) \p_y^n e^{ \ii \tau |y|} +
    \Oh(\lambda^2 \tau^n e^{-\hf |x| + \hf \ii \tau |y|})
    + \Oh(\lambda^2 e^{-\hf (|x| + |y|)}), \\
    \p_y^n f_{11} &= 2 \ii \tau \zeroth U^+_1(x) \p_y^n \zeroth Z^+_1(
    |y|)
    + \Oh(\lambda^2 e^{-\hf (|x| + |y|)}), \\
    \p_y^n f_{21} &=
    \begin{aligned}[t]
      -(k^2 + \tau^2 + \Oh(\lambda^3)) \big(
      & e^{\ii \tau |x|} \p_y^n \zeroth Z^+_1(|y|) \\
      &+ \Oh(e^{-\hf (|x| + |y|)}) + \Oh(\lambda^2 e^{\hf \ii \tau |x| - \hf
        |y|}) \big),
    \end{aligned} \\
    \p_y^n f_{22} &= \Oh(\lambda^2 \tau^n e^{\hf \ii \tau (|x| + |y|)}).
  \end{aligned}
\end{equation}
When $0 < y < x$ equation \eqref{eq:U+Z+} produces equations
\eqref{eq:f-common} plus the following terms:
\begin{align*}
  \p_y^n f_{13} &=
  \Oh( \lambda^2 e^{-\hf |x - y|}), \\
  \p_y^n f_{23} &=
  \begin{aligned}[t]
    -(k^2 + \tau^2 + \Oh(\lambda^3)) \bigg(
    & \p_y^n \frac{e^{\ii \tau (|x| + |y|)}
      - e^{\ii \tau |x - y|}}{2 \ii \tau} \\
    &+ \Oh(\lambda^2 \tau^{n-1} e^{\hf \ii \tau |x - y|})
    + \Oh(e^{-\hf |x - y|}) \\
    &+ \Oh(\lambda e^{\hf \ii \tau |x - y| - \hf |y|}) \bigg),
  \end{aligned} \\
  \p_y^n f_{14} &= \Oh(\lambda^2 e^{-\hf |x-y|}).
\end{align*}
When $y < x < 0$ equation \eqref{eq:U-Z-} gives the terms of
\eqref{eq:f-common} except for a sign change in $f_{21}$ and
\begin{align*}
  \p_y^n f_{31} &=
  \Oh((|k^2 + \tau^2| + |\lambda|^3) e^{-\hf|x-y|}), \\
  \p_y^n f_{32} &=
  \begin{aligned}[t]
    -(k^2 + \tau^2 + \Oh(\lambda^3)) \bigg(
    & \p_y^n \frac{e^{\ii \tau (|x| + |y|)}
      - e^{\ii \tau |x - y|}}{2 \ii \tau} \\
    &+ \Oh(\lambda^2 \tau^{n-1} e^{\hf \ii \tau |x - y|})
    + \Oh(\lambda^2 e^{-\hf |x - y|}) \\
    &+ \Oh(\tau^n e^{\hf \ii \tau |x - y| - \hf |x|}) \bigg),
  \end{aligned} \\
  \p_y^n f_{41} &=
  \Oh((|k^2 + \tau^2| + |\lambda|^3) e^{-\hf|x-y|}).
\end{align*}
All the other $f_{ij}$ are zero.  The case $x < y$ reduces to the
preceding cases by $F(x,y) = F(-x,-y)$.  We treat the leading terms
separately and summarize:
\begin{theorem} \label{th:F}
  There exists $\epsilon > 0$ such that for $|\lambda| < \epsilon$ the
  kernel $F$, where $F := (\det \Omega) R = F_0 + F_1$
  and $F_1 = F_{10} + F_{11} + F_{12}$, satisfies
  \begin{align*}
    F_0( x, {y}) &= \zeroth U^+_1(x)
    (e^{\ii \tau |{y}|} + 2 \ii \tau \zeroth Z^+_1(|{y}|)), \\
    F_{10}( x, {y}) &=
    \begin{cases}
      -\frac{k^2 + \tau^2}{2 \ii \tau}
      (e^{\ii\tau |x + {y}|} - e^{\ii\tau |x - {y}|})
      &\text{when $x {y} > 0$,} \\
      0 &\text{when $x {y} < 0$,}
    \end{cases} \\
    F_{11}( x, {y}) &=
    -{\rm sgn}( x(x - {y})) (k^2 + \tau^2)
    e^{\ii \tau |x|} \zeroth Z^+_1(|{y}|) \\
    |\partial_{y}^n F_{12}( x, {y})|
    &< C \bigl( |\lambda^3 \tau^{n-1} e^{\hf \ii \tau |x - {y}|}|
    + |\lambda^2 e^{-\hf |x - {y}|}|
    + |\lambda^3 e^{\hf \ii \tau |x| - {\frac{1 }{4}}|{y}|}| \bigr).
  \end{align*}
\end{theorem}
$F_0$ has two continuous derivatives with respect to ${y}$.  Although
we have tried to write these expressions so that they would be valid
everywhere the other pieces are a bit rough around the edges.  Hence
the estimate for the derivatives is only valid for $x \ne {y} \ne 0
\ne x$.  $F$ itself has two continuous derivatives, see the remark
after Theorem~\ref{th:reskernel}.

We finish this section with an estimate
for $R := (\zeta - D_k H_k)^{-1}$ when $|\zeta|$ is
large:
\begin{theorem} \label{th:respert}
  Let $k < \epsilon$.  For any $\alpha \in (0,\frac\pi4)$ there exist
  $C, c, r > 0$ such that for all $\tau$ with $|\tau| > r$, $\arg \tau
  \in (\alpha,\pi-\alpha)$ and $n \in \{0,1\}$
  \[
  |(R D_k^n)(x,{y})|
  < \frac C{|\tau|^{\frac32 - n}} e^{-c \sqrt{|\tau|} |x - {y}|}.
  \]
\end{theorem}
\begin{proof}
  We use the second Neumann series
  \begin{equation} \label{eq:Neumann}
    (\zeta - D_k H_k)^{-1} = (\zeta - A - B)^{-1}
    = \sum_{j=0}^\infty ((\zeta - A)^{-1} B)^j (\zeta - A)^{-1}
  \end{equation}
  with $A := D_k^2 + D_k$ and $B := D_k V$.  Let $R_\infty := (\zeta -
  A)^{-1}$.  $|V(x)| \le 6 e^{-|x|}$ but we also need exponential
  estimates for the kernels of $R_\infty$ and $R_\infty D_k$.  It is
  easier to work with their Fourier transforms.  The poles of $\hat
  R_\infty$ are $\ii\mu_j$ where the $\mu_j$ are given by
  \eqref{eq:muoftau}.  Expanding in partial fractions we have
  \begin{align*}
    \hat R_\infty(p) &= \frac1{\mu_1^2 - \mu_2^2} \left(
      \frac1{p^2 + \mu_1^2} - \frac1{p^2 + \mu_2^2}
    \right) \\
    \widehat{ R_\infty D_k}(p) &= \frac1{\mu_1^2 - \mu_2^2} \left(
      \frac{k^2 - \mu_1^2}{p^2 + \mu_1^2}
      - \frac{k^2 - \mu_2^2}{p^2 + \mu_2^2}
    \right).
  \end{align*}
  When $|\tau|$ is large $µ_j = \Oh(\sqrt\tau)$ and by \eqref{eq:murel}
  the coefficients of the partial fraction expansions are $\Oh(1/\tau)$
  and $\Oh(1)$ respectively.  Assume $|\Re \mu_j| > 2 a$ for some $a$.
  Then
  \[
  \int_{\bbR + \ii a} \frac1{|p^2 + \mu_j^2|} \,\frac{\dee p}{2\pi}
  < \frac\pi a.
  \]
  Thus $|R_\infty D_k( x, y)| \le \frac Ca e^{-a |x - y|}$.  If $\tau$
  is in an appropriate sector away from the real axis we can choose
  $a$ proportional to $\sqrt\tau$ and get convergence for the Neumann
  series.  The estimate of the theorem then follows easily.  For
  slightly more details see~\cite{timo}.
\end{proof}

\section{Estimates for the semigroup}

To complete our paper we are left with the task of
deriving the semigroup estimates of Section~\ref{linear}, mainly from the
resolvent estimates of Section~\ref{keres}.
However, when $t$ is small or $k$ large a crude estimate,
following from standard Fourier analysis and perturbation theory, will
suffice:
\begin{theorem} \label{th:sgpert}
There exist $C,c > 0$ such that for any $a \in (0,1]$
  \begin{align*}
    |e^{-t D_k H_k}(x,{y})|
    &< \frac C{t^{1/4}} e^{(\alpha(k,a) + c) t - a |x-{y}|} \\
    |(e^{-t D_k H_k} D_k)(x,{y})|
    &< \frac C{t^{3/4}} e^{(\alpha(k,a) + c) t - a |x-{y}|}
  \end{align*}
  where $t > 0$, $x,y \in \bbR$, and 
  \[
  \alpha(k,a) := -\tfrac78 (k^4 + k^2) + c (a^4 + a^2).
  \]
\end{theorem}
Notice that for bounded $t$ we can choose $a = t^{-1/4}$ to get rapid
decrease in $|x-{y}|$, or for large $k$ we can choose $a = \epsilon k$
for an $\epsilon$ such that $\alpha(k,a) + c < -{\frac{3 }{4}} k^4$.
\begin{proof}
  Another perturbation theory argument with $A = D_k^2 + D_k$ and $B =
  D_k V$.  This time use the expansion
  \begin{align*}
    e^{-t D_k H_k} &= \sum_{n=0}^\infty T_n(t) \\
    T_n(t) &= (-1)^n \idotsint\limits_{\{0 \le t_1 \le \cdots \le t_n \le t\}}
    e^{-(t-t_n)A} B e^{-(t_n - t_{n-1})A} B \cdots e^{-t_1 A}
    \,\dee t_1 \cdots \dee t_n.
  \end{align*}
  Fourier transform and the usual imaginary translation trick yield
  \begin{align*}
    |(e^{-t A} D_k^j)(x,\xi)| &< \frac C{t^{\fh + \frac j2}}
    e^{\alpha(k,a) t - a |x-\xi|} \\
    |(e^{-t A} B)(x,\xi)| &< \frac M{t^{3/4}}
    e^{\alpha(k,a) t - a |x-\xi| - |\xi|}
  \end{align*}
  with some $C, c, M > 0$ valid for all $a > 0$, $j \in \{0,1\}$.  By
  induction we get
  \begin{align*}
    |T_n(t;x,\xi)| &\le
    \frac{C (2 M \Gamma(\frac14))^n}{\Gamma(\frac{n+3}4)} t^{\frac{n-1}4}
    e^{\alpha(k,a) t - a |x-\xi|}, \\
    |(T_n D_k)(t;x,\xi)| &\le
    \frac{C (2 M \Gamma(\frac14))^n}{\Gamma(\frac{n+1}4)} t^{\frac{n-3}4}
    e^{\alpha(k,a) t - a |x-\xi|}.
  \end{align*}
  The prefactors can be estimated by the power series of $e^{c t}$,
  yielding the theorem.  For more details, see \cite{timo}.
\end{proof}

For other parameter values we use the Dunford--Cauchy integral
\eqref{reso} and the resolvent estimates.  There we have more cases:
for small $k$ and $\zeta$ we use Theorem~\ref{th:F}, for large $\zeta$
Theorem~\ref{th:respert} and for other $k$ and $\zeta$ just
Theorem~\ref{th:reskernel}.

Let $\epsilon >0$ be as in Section~\ref{keres} and Theorem~\ref{th:F},
assume $0 < k < \epsilon$ and that $t$ is bounded away from 0.  There
is a spectral gap between the lowest eigenvalue at $\Oh(k^3)$ and the
rest of the spectrum from $\Oh(k^2)$ upwards. By analyticity, the
integration path in \eqref{reso} can be modified to consist of two
parts: one surrounding the smallest eigenvalue and yielding a term
which we denote by $\pole K$; another part surrounding the rest of the
spectrum and yielding a term $\rest K$.

We first estimate $\pole K$.  Changing the integration variable to the
$\tau$ of \eqref{eq:zetaoftau} and using the notation of
Section~\ref{keres} the integral becomes
\begin{equation} \label{eq:tausgint}
  \int e^{-\zeta(k,\tau) t} \frac{2 \tau (1 + 2 k^2)^2}{\det \Omega}
  F( k, \tau; x, {y}) \, \deeC{\tau}.
\end{equation}
The integration path can be taken to run in the upper half plane.
Near the origin there is a pole, corresponding to the smallest
eigenvalue of $D_k H_k$.  To circle the spectrum the integration path
would need to pass above the pole but we integrate below the
pole instead and add the residue $\pole K$, which we now estimate.

We see from \eqref{eq:detOmega} that the zero of $\det \Omega$ is at
\[
p(k) := \ii k - \frac{\ii k^2}6 + \Oh(k^3).
\]
We have $\det \Omega = (\tau - p(k)) (2 \ii k + \Oh(\tau -
p(k)) + \Oh(k^2))$ and
\begin{equation} \label{eq:resdet}
  \lim_{\tau \to p(k)} \frac{2 \tau (\tau - p(k))}{\det \Omega}
  = \frac{p(k)}{\ii k + \Oh(k^2)} = 1 + \Oh(k).
\end{equation}
$p(k)$ corresponds to $\zeta_0(k) = \tfrac13 k^3 + \Oh(k^4)$.  The terms
of Theorem~\ref{th:F} produce $\pole K = \pole K_0 +
\pole K_1$ with
\begin{align}
  \pole K_0
  &= (e^{-\frac13 k^3 t} + \Oh(k e^{-\frac7{24} k^3 t})) \zeroth U^+_1(x)
  (e^{- (k + \Oh(k^2)) |{y}|} - 2 k \zeroth Z^+_1(|{y}|)),
  \label{eq:Kpole0}\\
  |\partial_{y}^n \pole K_1|
  &<
  \begin{aligned}[t]
    C e^{-{\frac{1 }{4}} k^3 t} \bigl(
    & |k^{2 + n} e^{- {\frac{1 }{4}} k |x - {y}|}| \\
    &+ |k^2 e^{-\hf |x - {y}|}|
    + |k^3 e^{- {\frac{1 }{4}} k |x| - {\frac{1 }{4}} |{y}|}| \bigr).
  \end{aligned} \label{eq:Kpole1}
\end{align}
where the $\Oh(\cdot )$ term comes from \eqref{eq:resdet} and replacing
$\zeta_0(k)$ by $\frac13 k^3$ (the number $\frac7{24}$
is just something between ${\frac{1 }{4}}$ and $\frac13$).

\begin{figure}
  \centering
  \setlength{\unitlength}{4144sp}%
\begingroup\makeatletter\ifx\SetFigFont\undefined%
\gdef\SetFigFont#1#2#3#4#5{%
  \reset@font\fontsize{#1}{#2pt}%
  \fontfamily{#3}\fontseries{#4}\fontshape{#5}%
  \selectfont}%
\fi\endgroup%
\begin{picture}(2274,744)(-11,107)
\thinlines
{\color[rgb]{0.285,0.301,0.285}\put(1126,839){\line( 0,-1){720}}
}%
{\color[rgb]{0.285,0.301,0.285}\put(  1,209){\line( 1, 0){2250}}
}%
{\color[rgb]{0.285,0.301,0.285}\put(1081,659){\line( 1, 0){ 90}}
}%
\put(1216,614){\makebox(0,0)[lb]{\smash{\SetFigFont{12}{14.4}{\familydefault}{\mddefault}{\updefault}{\color[rgb]{0.285,0.301,0.285}$\ii k$}%
}}}
{\color[rgb]{0,0,0}\put(2165,633){\line(-4,-1){796}}
\put(1369,434){\line(-1, 0){450}}
\put(919,434){\vector(-4, 1){832}}
}%
\end{picture}
  \caption{The integration path for $\rest K$.}
  \label{fig:taugamma}
\end{figure}
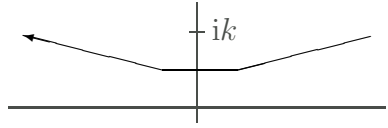

To estimate $\rest K$, the integral around the rest of the spectrum,
we again change to $\tau$ of \eqref{eq:zetaoftau}.  This leads to
integrating above the real axis but below the pole at $\approx \ii k$
as in Figure~\ref{fig:taugamma}.
We stay far enough
from the pole so that $\frac{|\tau|^2 + k^2}{k^2 + \tau^2}$ remains
bounded.  This allows us to write for small $\tau$
\[
\frac{2\tau}{\det\Omega} = \frac{2\tau}{k^2 + \tau^2} + \Oh(1)
\]
c.f.~\eqref{eq:detOmega}.  Denote $\tilde R :=
\frac{\dee\zeta}{\dee\tau} \, (\zeta(k,\tau) - D_k H_k)^{-1}$.
Combining with Theorem~\ref{th:F} we split $\tilde R$ as follows:
\begin{align*}
  \tilde R &= \tilde R_{00} + \tilde R_{01} + \tilde R_{10}
  + \tilde R_{11} + \tilde R_{12} \\
  \tilde R_{00}(x,{y}) &= \frac{2 \tau}{k^2 + \tau^2} \zeroth U^+_1(x)
  (e^{\ii \tau |{y}|} + 2 \ii \tau \zeroth Z^+_1(|{y}|)), \\
  \tilde R_{01}(x,{y}) &= \omega(k, \tau) \zeroth U^+_1(x)
  (e^{\ii \tau |{y}|} + 2 \ii \tau \zeroth Z^+_1(|{y}|)), \\
  \tilde R_{10}(x,{y}) &=
  \begin{cases}
    \ii (e^{\ii\tau |x + {y}|} - e^{\ii\tau |x - {y}|})
    &\text{when $x {y} > 0$,} \\
    0 &\text{when $x {y} < 0$,}
  \end{cases} \\
  \tilde R_{11}( x, {y}) &=
  -2 \tau {\rm sgn}( x(x - {y}))
  e^{\ii \tau |x|} \zeroth Z^+_1(|{y}|) \\
  |(-\partial_{y}^2 + k^2)^n \tilde R_{12}( x, {y})|
  &< \begin{aligned}[t]
    C \bigl( & |\lambda^{2n+1} e^{-\mu(\tau) |x - {y}|}| \\
    &+ |\lambda e^{-\hf |x - {y}|}|
    + |\lambda^2 e^{\hf \ii \tau |x| - {\frac{1 }{4}}|{y}|}| \bigr)
  \end{aligned}
\end{align*}
where $\omega$ is bounded, $\mu(\tau) := \min\{c \Im \tau, 1\}$ for some
$c > 0$ and $n \in \{0,1\}$.  We derived this for small $\tau$ but by
Theorem~\ref{th:respert} it actually holds also for large $\tau$ on
our integration path.  Between small and large $\tau$ we have
Theorems~\ref{th:Omegainv}, \ref{th:reskernel} and
Lemma~\ref{lm:yzopposite}.  Everything is continuous on a compact
interval, hence bounded.  By choosing $c$ appropriately $-\mu(\tau)$ can
be used to estimate $\Re \mu_2$ from above and thus our estimate holds
everywhere along the integration path.

As the explicit terms are analytic we can simply integrate them along
the real axis ($e^{-\zeta(k,\tau) t}$ makes everything small as
$|\tau| \to \infty$).  Denote $t' := (1 + 2 k^2)^2 t$.
\begin{align}
  \rest K_{00} &= -\zeroth U^+_1(x) e^{-(k^2 + k^4) t}
  \int_{-\infty}^\infty
  \frac{2 \tau}{k^2 + \tau^2}  e^{-\tau^2 t'} \left(
    e^{\ii \tau |{y}|} + 2 \ii \tau \zeroth Z^+_1(|{y}|) \right)
  \,{ \deeC{\tau} } \notag\\
  & \begin{aligned}
    = e^{(3 k^4 + 4 k^6) t} \zeroth U^+_1(x) \Biggl(
    & f( k \sqrt{t'}, \tfrac{|{y}|}{2 \sqrt{t'}})
    - f( k \sqrt{t'}, -\tfrac{|{y}|}{2 \sqrt{t'}}) \\
    &+ \biggl(
    4 k \Gtail( k \sqrt{t'}) - \frac{2 e^{-k^2 t'}}{\sqrt{\pi t'}}
    \biggr) \zeroth Z^+_1(|{y}|)
    \Biggr)
  \end{aligned} \label{eq:Krest00}
\end{align}
where
\[
\Gtail(x) := \frac1{\sqrt\pi} \int_x^\infty e^{-r^2} \,\dee r
\qquad \text{and} \qquad
f( x, y) := e^{2 x y} \Gtail( x + y).
\]
The coefficient of the $\zeroth Z^+_1$ was obtained from the
requirement that $\rest K_{00}(x,{y})$ be continuously
differentiable in ${y}$ (because $\tilde R_{00}$ is).
\begin{align}
  \rest K_{10} &= -e^{-(k^2 + k^4) t}
  \int_{-\infty}^\infty e^{-\tau^2 t'}
  \bigl( e^{\ii \tau |x + {y}|} - e^{\ii \tau |x - {y}|} \bigr) \,
  {\frac{ \dee\tau }{2\pi  }} \notag\\
  &= \frac{e^{-(k^2 + k^4) t}}{\sqrt{4 \pi t'}}
  \Bigl( e^{-\frac{(x-{y})^2}{4 t'}} - e^{-\frac{(x+{y})^2}{4 t'}} \Bigr)
  \label{eq:Krest10}
\end{align}
when $x {y} > 0$ and zero otherwise.
\begin{align}
  \rest K_{11}
  &= e^{-(k^2 + k^4) t} {\rm sgn}( x (x - {y})) \zeroth Z^+_1(|{y}|)
  \int_{-\infty}^\infty 2 \tau e^{-\tau^2 t' + \ii \tau |x|}
  \,{ \deeC{\tau} } \notag\\
  &= {\rm sgn}( x - {y}) \frac x{\sqrt{4\pi} (t')^{\frac32}}
  e^{-(k^2 + k^4) t - \frac{x^2}{4 t'}}  \zeroth Z^+_1(|{y}|).
  \label{eq:Krest11}
\end{align}
The rest can be estimated by integrating so that $\Im \tau > \hf k$.
\begin{align}
  (-\partial_{y}^2 + k^2)^n \rest K_{01} &=
  e^{-\hf k^2 t} \zeroth U^+_1(x)
  \Oh(\frac1{t^{n+\hf}} e^{-\hf k |{y}|} + \frac1t e^{-|{y}|})
  \label{eq:Krest01}\\
  |(-\partial_{y}^2 + k^2)^n \rest K_{12}| &<
  \begin{aligned}[t]
    C e^{-\hf k^2 t} \biggl( & \frac1{t^{n + 1}}
    e^{-\frac c2 k |x - {y}|} \\
    &+ \frac1t e^{-\hf |x - {y}|} + \frac1{t^{\frac32}} e^{-\frac c2 k
      |x| - {\frac{1 }{4}} |{y}|} \biggr).
  \end{aligned} \label{eq:Krest12}
\end{align}

When $k < 1/\sqrt t$ (in addition to $k < \epsilon$) a slightly
different estimate is useful: instead of treating the pole
separately we integrate around the whole spectrum at once.  Everything
goes essentially as in $(ii)$ except that now the
integration path goes above $\ii k$.  This affects the $\tilde R_{00}$
integral:
\begin{align}
  \all K_{00} &= -\zeroth U^+_1(x) e^{-(k^2 + k^4) t}
  \int_{-\infty + \frac{2i}{\sqrt t}}^{\infty + \frac{2i}{\sqrt t}}
  \frac{2 \tau}{k^2 + \tau^2}  e^{-\tau^2 t'} \left(
    e^{\ii \tau |{y}|} + 2 \ii \tau \zeroth Z^+_1(|{y}|) \right)
  \,{\deeC{\tau}} \notag\\
  &= \begin{aligned}[t]
    e^{(3 k^4 + 4 k^6) t} \zeroth U^+_1(x) \biggl(
    & f( k \sqrt{t'}, \tfrac{|{y}|}{2 \sqrt{t'}})
    + f( -k \sqrt{t'}, \tfrac{|{y}|}{2 \sqrt{t'}}) \\
    &+ \Bigl(
    4 k \Gtail( k \sqrt{t'}) - 2 k
    - \frac{2 e^{-k^2 t'}}{\sqrt{\pi t'}}
    \Bigr) \zeroth Z^+_1(|{y}|)
    \biggr)
  \end{aligned} \label{eq:Kall00}
\end{align}
and the estimates for the remainders:
\begin{align}
  (-\partial_{y}^2 + k^2)^n \all K_{01} &= \zeroth U^+_1(x)
  \Oh(\frac1{t^{n+\hf}} e^{-\frac{2|{y}|}{\sqrt t}} + \frac1t e^{-|{y}|}),
  \label{eq:Kall01}\\
  |(-\partial_{y}^2 + k^2)^n \all K_{12}|
  &< \begin{aligned}[t]
    C \biggl( & \frac1{t^{n + 1}} e^{-\frac{2c|x - {y}|}{\sqrt t}} \\
    &+ \frac1t e^{-\hf |x - {y}|} + \frac1{t^{\frac32}}
    e^{-\frac{2c|x|}{\sqrt t} - {\frac{1 }{4}} |{y}|} \biggr)
  \end{aligned} \label{eq:Kall12}
\end{align}
where $C$ depends on $c$.  The other two terms are not effected as
they are analytic in $\tau$.

\section{Proof of Lemmas \protect\ref{lm:K-smallk} and
  \protect\ref{lm:S-smallk}}
\label{sec:sgsummary}

The above calculations contain more than enough to prove Lemmas
\ref{lm:K-smallk} and \ref{lm:S-smallk}.  Take $k_0$ equal to the
$\epsilon$ of Section~\ref{keres}.  Let us first treat $K_0$ in the
case $k \le 1/\sqrt t$.  Then define $K_0 := \all K_{00} + \all
K_{01}$.  \eqref{eq:Kall00} and \eqref{eq:Kall01} yield \eqref{timo1} and
\[
\all K_0 = \zeroth U^+_1(x) (1 + \Oh(\frac1{\sqrt t}(1 + |{y}|))),
\]
whence it follows for $h \in X$ that
\[
\begin{split}
  \int_{\bbR^d} e^{-\ii \svec{k} \cdot \svec{y}}
  \all K_0 h \,\dee\bvec{y}
  &= \zeroth U^+_1(x) \Bigl(
  \int e^{-\ii \svec k \cdot \svec y} h \,\dee\bvec y
  + \Oh(\frac{\|h\|_X}{\sqrt t}) \Bigr) \\
  &= \zeroth U^+_1(x) \Bigl(
  \int (1 + \Oh( \svec k \cdot \svec y)) h \,\dee\bvec y
  + \Oh(\frac{\|h\|_X}{\sqrt t}) \Bigr) \\
  &= \zeroth U^+_1(x) \Bigl(
  \int h \,\dee\bvec{y} + \Oh(\frac{\|h\|_X}{\sqrt t}) \Bigr).
\end{split}
\]
Hence we have established \eqref{zprop} for $k \le 1/\sqrt t$.

When $1/\sqrt t < k \le k_0$ define $K_0 := \pole K_0 + \rest K_0$ and
$\rest K_0 := \rest K_{00} + \rest K_{01}$.  Using \eqref{eq:Kpole0},
\eqref{eq:Krest00} and \eqref{eq:Krest01} from the
previous section we get \eqref{timo3} and
\begin{align}
  \pole K_0 &= \zeroth U^+_1(x)
  (e^{-\frac13 k^3 t} + \Oh(k(1 + |{y}|) e^{-\frac7{24} k^3 t}),
  \label{kaava1} \\
  \rest K_0 &= \zeroth U^+_1(x)
  \Oh(\frac1{\sqrt t}(1 + |{y}|) e^{-\hf k^2 t}).
  \label{kaava2}
\end{align}
Similarly to the previous case we get
\begin{align*}
  \int_{\bbR^d} e^{-\ii \bvec{k} \cdot \bvec{y}}
  \pole K_0 h \,\dee\bvec{y}
  &= \zeroth U^+_1(x) \Bigl(
  e^{-\frac13 k^3 t} \int h \,\dee\bvec{y}
  + \Oh(\frac{\|h\|_X}{t^{1/3}} e^{-{\frac{1 }{4}} k^3 t}) \Bigr), \\
  \int_{\bbR^d} e^{-\ii \bvec{k} \cdot \bvec{y}}
  \rest K_0 h \,\dee\bvec{y}
  &= \zeroth U^+_1(x) \Oh(\frac{\|h\|_X}{\sqrt t} e^{-\hf k^2 t})
\end{align*}
and \eqref{zprop} follows.

Let us then proceed with $K_1$.  Starting again with $k \le 1/\sqrt t$
write $K_1 := \rest K_{10} + \rest K_{11} + \all K_{12}$.  $\rest
K_{10}$, given by \eqref{eq:Krest10}, is essentially the first term of
\eqref{timo2-}; the difference can be absorbed into $K_2$ together
with the other two terms, for which see \eqref{eq:Krest11} and
\eqref{eq:Krest12}.  Thus \eqref{timo2--} is satisfied.  \eqref{timo2}
is also satisfied.  When $1/\sqrt t < k \le k_0$ define $K_1 := \pole
K_1 + \rest K_{10} + \rest K_{11} + \all K_{12}$.  Using
\eqref{eq:Kpole1}, \eqref{eq:Krest10}, \eqref{eq:Krest11} and
\eqref{eq:Kall12} it is easy to see that \eqref{timo2--} and
\eqref{timo4} are satisfied.

Finally, 
Lemma~\ref{bigk} (b) follows from Theorem~\ref{th:sgpert} and
Lemma~\ref{bigk} (a) follows from the following theorem:
\begin{theorem} \label{th:sg-non-small-k}
  For any $\epsilon > 0$ there exist $C, c > 0$ such that for any $k >
  \epsilon$, $t > 1$ and $n \in \{0,1\}$
  \[
  |(e^{-t D_k H_k} D_k^n)(x,{y})| < C e^{-\hf k^4 t - c |x-{y}|}
  \]
\end{theorem}
\begin{proof}
  Theorem~\ref{th:sgpert} takes care of things for very large $k$, say
  $k > r$.  For $k \in [\epsilon, r]$ we use the Dunford--Cauchy
  integral
  \[
  e^{-t D_k H_k} = \int_\Gamma e^{-\zeta t}(\zeta - D_k H_k)^{-1}
  \,{\deeC{\zeta}}
  \]
  and estimate the resolvent.  The spectrum has $k^4$ as a lower bound
  by Lemma~\ref{lm:coarse-spectrum}.  Thus $\Gamma$ can be chosen so
  that $\Re \zeta > \hf k^4$ and $\Re \mu_1 \le \Re \mu_2 < -c < 0$
  for some $c$.  Asymptotically the path can be chosen to be $s
  \mapsto \hf k^4 + s e^{\hf \ii \alpha}$ for some $\alpha < \pi/2$.
  The resolvent kernel is then bounded by $C e^{-c |x - {y}|}$: for
  large $|\zeta|$ this follows from Theorem~\ref{th:respert},
  otherwise just from \eqref{eq:reskernel} and continuity.
\end{proof}

\end{document}